%% file: main.tex
\pdfoutput=1
\documentclass[sigconf,natbib=true,anonymous=false]{acmart}

\AtBeginDocument{%
  \providecommand\BibTeX{{%
    \normalfont B\kern-0.5em{\scshape i\kern-0.25em b}\kern-0.8em\TeX}}}

\copyrightyear{2026}
\acmYear{2026}
\setcopyright{cc}
\setcctype{by}
\acmConference[SIGIR '26]{Proceedings of the 49th International ACM SIGIR Conference on Research and Development in Information Retrieval}{July 20--24, 2026}{Melbourne, VIC, Australia}
\acmBooktitle{Proceedings of the 49th International ACM SIGIR Conference on Research and Development in Information Retrieval (SIGIR '26), July 20--24, 2026, Melbourne, VIC, Australia}
\acmDOI{10.1145/3805712.3809756}
\acmISBN{979-8-4007-2599-9/2026/07}

\usepackage{placeins} 
\usepackage{tcolorbox}       
\usepackage{listings}        
\usepackage{soul}

\usepackage{balance}
\usepackage{pifont}
\usepackage{amsfonts, amsthm, amsmath}
\usepackage[ruled,vlined,linesnumbered]{algorithm2e}
\usepackage{color, colortbl}
\usepackage{enumitem,multirow,graphicx,subcaption,multicol,lipsum,float,adjustbox}
\newlength{\textfloatsepsave}  
\usepackage{cleveref}
\usepackage{algorithmicx}
\usepackage[page]{appendix} 
\crefformat{section}{\S#2#1#3}
\crefformat{subsection}{\S#2#1#3}
\crefformat{subsubsection}{\S#2#1#3}
\usepackage[normalem]{ulem}
\useunder{\uline}{\ul}{}

\begin{document}


\title{SPRINT: Scalable and Predictive Intent Refinement for LLM-Enhanced Session-based Recommendation}

\begin{abstract}
\input{./sections/000abstract}
\end{abstract}

\ccsdesc[500]{Information systems~Recommender systems}

\vspace{-1cm}
\keywords{Session-based Recommendation, LLM-based user profiling}
\input{dfn}

\author{Gyuseok Lee}
\affiliation{
    \institution{University of Illinois Urbana-Champaign	}
    \city{Champaign}
    \state{IL}
    \country{USA}
}
\email{gyuseok2@illinois.edu}

\author{Wonbin Kweon}
\affiliation{
    \institution{University of Illinois Urbana-Champaign	}
    \city{Champaign}
    \state{IL}
    \country{USA}
}
\email{wonbin@illinois.edu}

\author{Zhenrui Yue}
\affiliation{
    \institution{University of Illinois Urbana-Champaign	}
    \city{Champaign}
    \state{IL}
    \country{USA}
}
\email{zhenrui3@illinois.edu}

\author{Yaokun Liu}
\affiliation{
    \institution{University of Illinois Urbana-Champaign	}
    \city{Champaign}
    \state{IL}
    \country{USA}
}
\email{yaokunl2@illinois.edu}

\author{Yifan Liu}
\affiliation{
    \institution{University of Illinois Urbana-Champaign	}
    \city{Champaign}
    \state{IL}
    \country{USA}
}
\email{yifan40@illinois.edu}

\author{Susik Yoon}
\affiliation{
    \city{}
    \institution{Korea University}
    \city{Seoul}
    \country{Republic of Korea}
}
\email{susik@korea.ac.kr}

\author{Dong Wang}
\affiliation{
    \institution{University of Illinois Urbana-Champaign	}
    \city{Champaign}
    \state{IL}
    \country{USA}
}
\email{dwang24@illinois.edu}

\author{SeongKu Kang}
\affiliation{
    \city{}
    \institution{Korea University}
    \city{Seoul}
    \country{Republic of Korea}
}
\authornote{Corresponding author.}
\email{seongkukang@korea.ac.kr}


\maketitle
\section{Introduction}

\input{./sections/010introduction.tex}
\section{Related Work}
\label{sec:relatedwork}

\input{./sections/060Related_work}

\section{Problem Formulation}
\label{sec:preliminary}
\input{./sections/020Preliminary}
\section{\proposed}
\label{sec:method}
\input{./sections/040Method}

\section{Experiments}
\subsection{Experimental Setup}
\label{sec:experimentsetup}
\input{./sections/050Experiment_setup}

\input{./sections/051Experiment_result}

\vspace{0.05cm}
\section{Conclusion}
\label{sec:conclusion}
\input{sections/070conclusion}

\vspace{0.05cm}
\section*{Acknowledgments}
This work was the result of project supported by KT(Korea Telecom)-Korea University AICT R\&D Center.
This work was also supported by ICT Creative Consilience Program through the IITP grant funded by the MSIT (IITP-2026-RS-2020-II201819), the NRF grant funded by the MSIT (RS-2026-25486220), and Basic Science Research Program through the NRF funded by the Ministry of Education (NRF-2021R1A6A1A03045425), and the IITP grant funded by the MSIT (IITP-2026-RS-2025-02304828).





\bibliographystyle{ACM-Reference-Format}
\balance
\bibliography{acmart}


\end{document}

%% file: sections/000Abstract.tex

Large language models (LLMs) have enhanced conventional recommendation models via user profiling, which generates representative textual profiles from users’ historical interactions.
However, their direct application to session-based recommendation (SBR) remains challenging due to severe session context scarcity and poor scalability.
In this paper, we propose \proposed, a scalable SBR framework that incorporates reliable and informative intents while ensuring high efficiency in both training and inference.
\proposed constrains LLM-based profiling with a global intent pool and validates inferred intents based on recommendation performance to mitigate noise and hallucinations under limited context. 
To ensure scalability, LLMs are selectively invoked only for uncertain sessions during training, while a lightweight intent predictor generalizes intent prediction to all sessions without LLM dependency at inference time. 
Experiments on real-world datasets show that \proposed consistently outperforms state-of-the-art methods while providing more explainable recommendations.\footnote{Our code is available at \href{https://github.com/Gyu-Seok0/SPRINT_SIGIR26}{https://github.com/Gyu-Seok0/SPRINT\_SIGIR26}.}
\vspace{-0.1cm}

%% file: dfn.tex
\newcommand{\proposed}{SPRINT\xspace}
\newcommand{\tRS}{$RS^T$\xspace}
\newcommand{\sRS}{$RS^S$\xspace}

\newcommand{\smallsection}[1]{{\vspace{0.03in} \noindent \bf {#1}}}

%% file: sections/010introduction.tex
\begin{figure}[t]
  \centering
  \hspace*{-0.25cm}
  \includegraphics[width=1.05\columnwidth]{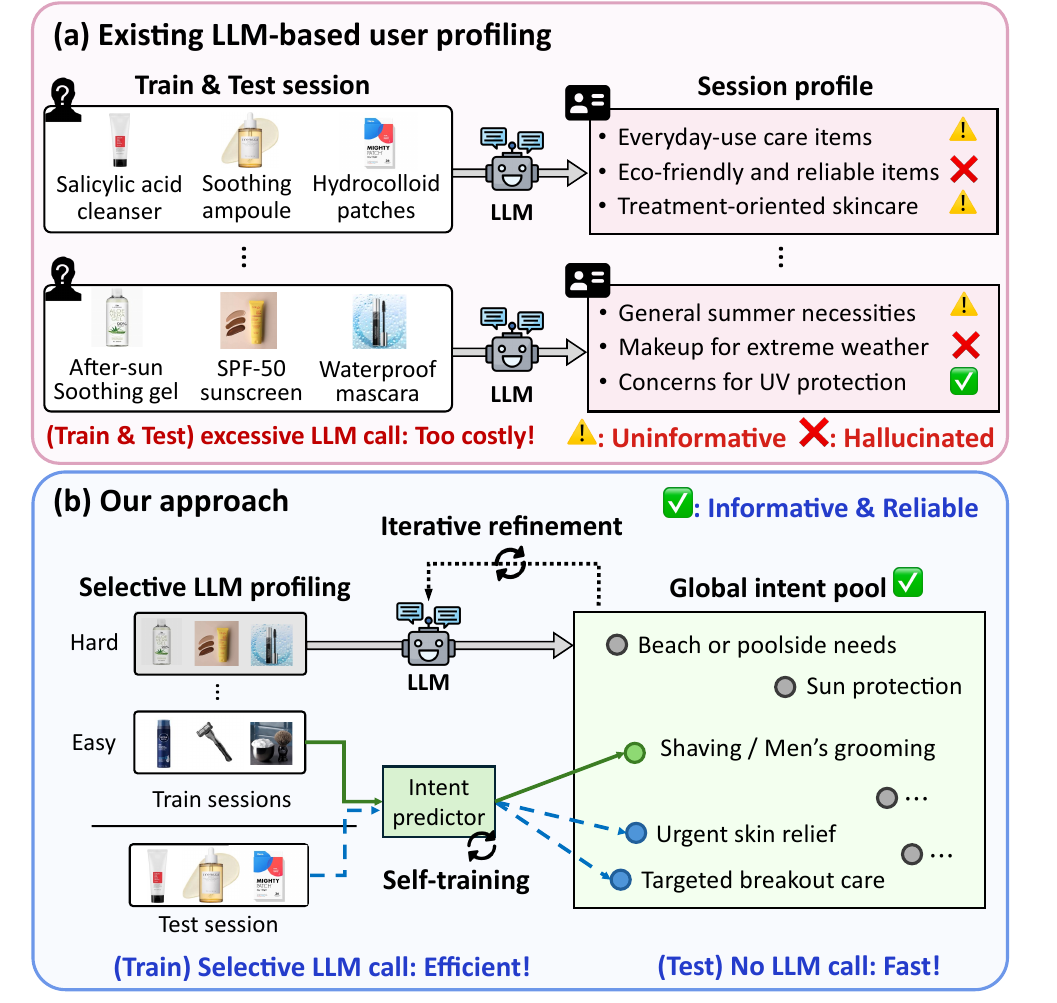}
  \vspace{-3mm}
  \caption{A conceptual comparison of (a) existing LLM-based user profiling and (b) our approach. Best~viewed~in~color.}
  \label{fig:concept}
\end{figure}


With extensive world knowledge and reasoning capabilities, large language models (LLMs) have significantly enhanced conventional recommendation models (CRMs)~\cite{zhao2024recommender, dai2023uncovering}.
Through a deep understanding of natural language (e.g., user reviews, item descriptions), LLMs provide high-quality semantic information that is not directly captured by raw user–item interaction data~\cite{wei2024llmrec, jia2025learn, kweon2025uncertainty, lee2025collaborative}.
Existing work has incorporated this knowledge through various strategies, including embedding transfer~\cite{liu2025llmemb, hu2024enhancing}, semantic feature augmentation~\cite{li2025ctrl, lyu2024llm}, and auxiliary supervision~\cite{wang2024rdrec, wang2025large}.
As a result, CRMs can leverage rich semantic knowledge to achieve substantially more accurate recommendations, while remaining highly efficient by avoiding costly LLM calls at inference time~\cite{wu2024survey, liu2025large}.


A prominent line of research is LLM-based \textit{user profiling}, which focuses on generating representative and descriptive textual profiles from users’ historical interaction data~\cite{RLMRec, LLMESR, ProfileRec, LLM4SBR}.
Since users’ preferences are only implicitly expressed through their behavioral histories, LLMs are leveraged for their knowledge and reasoning capabilities to unveil such latent preferences and synthesize them into explicit high-level descriptions, such as user profiles~\cite{RLMRec, ProfileRec}, personas~\cite{shi2025you, wang2022learning}, or intents~\cite{LLM4SBR, lian2025egrec}.
These profiles supply CRMs with explicit information about users’ underlying preferences that are challenging for models such as SASRec~\cite{sasrec} to uncover solely from behavioral sequences.

However, directly applying existing LLM-based user profiling to session-based recommendation (SBR) is non-trivial.
SBR is designed for practical scenarios, as many users browse online services without logging in or leaving persistent profiles~\cite{gru4rec, ProxySR}. 
These design constraints shape the unique characteristics of session data: it is typically short and anonymous, preventing the use of user priors (e.g., metadata and historical records).
Moreover, the absence of user IDs makes it impossible to determine which sessions belong to the same user, forcing each session to be processed independently and resulting in a massive volume of isolated sessions.
Consequently, directly applying LLM-based user profiling to SBR faces two critical challenges: (C1) context scarcity and (C2) scalability issues.

\noindent\textbf{(C1) Uninformative profiles due to session context scarcity}: 
Compared to traditional RS, SBR suffers from a more severe paucity of information~\cite{LLM4SBR, MiasRec}, which renders most existing LLM-based profiling methods ineffective.
Specifically, each session contains only a small number of interactions (e.g., the average session length of Yelp is 3.63~\cite{song2019session}), making it difficult to generate informative user profiles.
Moreover, the anonymous nature of sessions hinders the capture of user traits and long-term preferences~\cite{qiu2020exploiting, ferrato2023challenges}.
In this context-deficient environment, existing methods relying on substantial textual input and previous records are prone to noise and hallucinations, leading to misleading user behavior predictions.
While a recent profile-based method~\cite{ProfileRec} adopts self-reflection~\cite{shinn2023reflexion, ji2023towards} to ensure that profiles are grounded in the input information, it primarily targets logical consistency rather than capturing user preferences.
Consequently, the generated profiles do not necessarily guarantee recommendation utility.

\noindent\textbf{(C2) Scalability issues in SBR training and inference}:
Existing LLM-based profiling approaches suffer from severe scalability issues in both training and inference when applied to the SBR setting.
During training, generating profiles for a massive number of sessions requires excessive LLM calls, making the offline pipeline significantly time-consuming and costly. 
Furthermore, the anonymity of sessions forces the system to generate profiles on-the-fly with LLMs during inference. 
This is further exacerbated by methods with advanced reasoning mechanisms such as iterative self-reflection~\cite{ProfileRec}, memory-based reflection~\cite{bougie2025simuser} or prompt optimization~\cite{PO4ISR}, which may result in more than a doubling of training and inference time. %
Such substantial overhead renders existing approaches impractical for real-world SBR deployment, highlighting the need for a more scalable generation strategy.

In this paper, we propose \textbf{\proposed}, a framework that enables \textbf{\underline{S}}calable and \textbf{\underline{PR}}edictive \textbf{\underline{INT}}ent  refinement for LLM-enhanced SBR (Figure~\ref{fig:concept}).
Specifically, \proposed consists of two stages:

\noindent \textbf{(1)} 
In the first stage, we infer the underlying intents of sessions reliably and efficiently using LLMs.
Since unconstrained free-form generation is prone to noise and hallucinations, we reformulate intent generation as an \emph{intent identification} problem over an expandable global intent pool.
This pool serves as a shared and bounded intent space across sessions.
To populate the intent pool in a scalable manner, we \emph{selectively invoke} LLMs only for sessions where the SBR model exhibits high uncertainty, indicating insufficient contextual understanding.
For such ambiguous cases, the inferred intents are refined through a predict-and-correct loop that iteratively validates their reliability based on their contribution to recommendation~performance, ensuring their predictive utility.

\noindent \textbf{(2)} 
In the second stage, the generated intents are utilized to enhance the SBR model via an \textit{intent predictor}.
The intent predictor is a lightweight module that infers multiple intents per session without LLM dependency and can be flexibly integrated with existing SBR models.
It is trained via a self-training strategy.
Specifically, LLM-generated intents provide initial supervision, while confident predictions on sessions without LLM-derived intents are iteratively incorporated as pseudo-labels based on their interaction patterns.
To further improve prediction accuracy, we devise a \textit{collaborative intent enrichment} process, in which the predictor refines its predictions by leveraging neighboring sessions with similar~behavioral~patterns.




SPRINT systematically addresses the aforementioned challenges.
For \textbf{C1}, SPRINT mitigates context scarcity by constraining LLM generation, validating inferred intents, and enriching them through collaborative signals across similar sessions.
For \textbf{C2}, SPRINT resolves scalability issues through uncertainty-aware LLM invocation during training and lightweight, LLM-free intent prediction at inference time.
Our core contributions are as follows:
\begin{itemize}[leftmargin=*]\vspace{-\topsep}
    \item We identify two fundamental challenges in applying LLM-based user profiling to SBR that remain underexplored in the literature, highlighting the need for systematic investigation.
    
  
    \item 
    We propose \proposed, a scalable SBR framework that generates reliable and informative intents via selective LLM usage during training and enables lightweight, LLM-free intent prediction at inference time.

    \item We validate the effectiveness of \proposed with comprehensive experiments and provide in-depth analyses of each proposed component.
    \proposed consistently outperforms state‑of‑the‑art methods while achieving substantially higher efficiency.
\end{itemize}

%% file: sections/060Related_work.tex
\noindent
\textbf{Session-based recommendation (SBR).}
SBR aims to predict the next item given an anonymous session by modeling short‑term user behavior~\cite{FAPAT, liu2018stamp, yoo2024ensuring, yoo2025continual, yoo2025embracing} and is widely adopted for delivering timely recommendations without user authentication~\cite{liu2018stamp, kersbergen2022serenade, liu2020keywords}.
Early RNN‑based methods encode temporal dynamics via gated recurrent units~\cite{gru4rec,narm}.
Subsequent GNN-based methods model complex item transitions by linking items across sessions in a unified graph~\cite{SR-GNN, GCE-GNN},
and Transformer-based models leverage self‑attention to capture relationships between distant items within a session~\cite{GC-SAN, CoSAN}.
However, SBR models suffer from data sparsity due to the short and anonymous~nature~of~sessions~\cite{wang2021survey, ProxySR}.
With such limited context, conventional SBR models struggle to capture reliable sequential patterns, resulting in suboptimal recommendations.
To address this issue, prior work~\cite{yang2023based, FAPAT, ELCRec, ProxySR} has explored modeling strategies that enrich session representations beyond raw interaction sequences.

\smallsection{Intent-aware recommendation.}
Among these efforts, a prominent direction is to model user intent within a session~\cite{chen2022intent, wang2021learning, yang2023based, li2023multi, qin2024intent}.
Intent reveals the underlying preferences behind user choices and provides contextual signals beyond mere interaction logs.
As intent has no ground-truth labels, existing intent modeling can be broadly categorized into two groups: 
(1) explicit intent modeling, which utilizes side information (e.g., categories, titles) as intent labels~\cite{CAFE, FAPAT, LLM4SBR}, and (2) implicit intent modeling, which infers latent intents via unsupervised techniques like clustering~\cite{ELCRec, chen2022intent, qin2024intent}.
Despite these advances, explicit methods remain vulnerable to noisy or incomplete side information, while implicit methods suffer from limited explainability and reliance on heuristic clustering choices.
These limitations motivate the use of LLMs as a new paradigm in intent modeling, leveraging pretrained language knowledge to generate more robust and interpretable intents~\cite{PO4ISR, LLM4SBR, lian2025egrec, wang2025re2llm, wang2025intent}.

\smallsection{LLM-based SBR with intent profiling.}
Recent LLM-based SBR methods adapt user profiling to anonymous sessions through session-level intent profiling, and generally follow two paradigms:
(1) LLM-as-SBR models~\cite{PO4ISR,wang2025re2llm} and (2) LLM-enhanced SBR~\cite{LLM4SBR,lian2025egrec}.

In the first paradigm, LLMs serve as standalone SBR models that directly infer user intents and rank items based on the intents. 
PO4ISR~\cite{PO4ISR} optimizes prompts to extract user intents and then ranks candidates accordingly.
Re2LLM~\cite{wang2025re2llm} reinforces prompts by injecting hints to fix recommendation errors.
Though effective, directly deploying LLMs for recommendation inevitably incurs substantial inference-time overhead, making them less preferred for real-time applications
(see \cref{sub:efficiency_analysis} for a detailed analysis).

In contrast, LLM-enhanced SBR methods employ LLMs to generate user intents that guide conventional SBR models.
EGRec~\cite{lian2025egrec} jointly trains an LLM for intent generation with a recommender, while LLM4SBR~\cite{LLM4SBR} infers long- and short-term intents from item titles to enrich session representations.
Despite their effectiveness, EGRec requires costly fine-tuning and risks degrading the LLM’s pretrained knowledge, while LLM4SBR limits intent expressiveness by treating intents as simple item titles.
Moreover, both methods rely on LLM-generated outputs that are not explicitly optimized for recommendation performance.
This can introduce spurious patterns and potentially degrade recommendation accuracy. 
These limitations highlight the need for an LLM-enhanced SBR approach that explicitly validates generated intents while remaining efficient and scalable for deployment.

%% file: sections/020Preliminary.tex
\subsection{Preliminaries}

\noindent
\textbf{Notations.}
Let \(S = [i_1, i_2, \dots, i_{N}]\) be a session of $N$ items from the item set $\mathcal{I}$.
Given $S$, the goal of SBR is to predict the next item $i_\text{next} $.
The set of all sessions $\mathcal{S}$ is partitioned into disjoint training and test sets, denoted by $\mathcal{S}_\text{train}$ and $\mathcal{S}_\text{test}$,
such that $\mathcal{S}_\text{train} \cap \mathcal{S}_\text{test} = \emptyset$.
Let $\mathcal{F}$ be the set of feature fields for items (e.g., title, description, categories, reviews). 
For each item, we collect its features into a tuple $\mathbf{x}_i = (x_f)_{f \in \mathcal{F}}$, where $x_f$ denotes the raw data for field $f$.
Given a session $S$, we define the session feature matrix as $\mathbf{X}_S = [\mathbf{x}_{1}, \mathbf{x}_{2}, \dots, \mathbf{x}_{{N}}]$.

\smallsection{Conventional SBR.} 
Let $\mathrm{SBR}(\cdot)$ be a conventional SBR model (e.g., RNN-~\cite{gru4rec, narm} and Transformer-based~\cite{GC-SAN, CoSAN}).
Given a session $S$, it estimates the probability distribution for the next item as follows:
\begin{equation}\label{eq:SBR}
    \begin{aligned}
          P_\theta(i_\text{next}\!\mid\!S) \longleftarrow \mathrm{SBR}(S, \mathbf{X}_S),
    \end{aligned}
\end{equation}
where \(\theta\) denotes the parameters of the SBR model. 
The next item is predicted as: $\hat{i}_{\text{next}} = {\operatorname{argmax}_{j\in \mathcal{I}}}\,P_{\theta}\bigl(j\mid S\bigr)$.


\subsection{Problem Definition}
We study LLM-based intent profiling in the context of SBR, where LLMs are used to generate explicit session-level intents to enhance conventional SBR models.
Let $\mathbf{Z}_S = \mathrm{LLM}(\mathbf{X}_S)$ be the output from the LLM, which may take various forms such as profiles~\cite{RLMRec, ProfileRec}, personas~\cite{shi2025you, wang2022learning}, or intents~\cite{LLM4SBR, lian2025egrec}.
These signals are then provided alongside the session:
\begin{equation}
    \begin{aligned}
        P_\theta(i_\text{next}\mid S) \longleftarrow \mathrm{SBR}\bigl(S,\; \mathbf{X}_S, \;\mathbf{Z}_S\bigr).
    \end{aligned}
\end{equation}
$\mathbf{Z}_S$ offers rich contextual information not directly observable from~$S$, thereby enhancing the learning of SBR models.
Despite its promise, LLM-based profiling for SBR is limited by two aforementioned challenges that remain insufficiently explored in prior work~\cite{LLM4SBR, RLMRec, ProfileRec, LLMESR}: (C1) uninformative profiles caused by session context scarcity, and (C2) scalability issues in SBR training and inference.

%% file: sections/040Method.tex
\begin{figure*}[t]
    \centering    
    \includegraphics[scale=1.4]{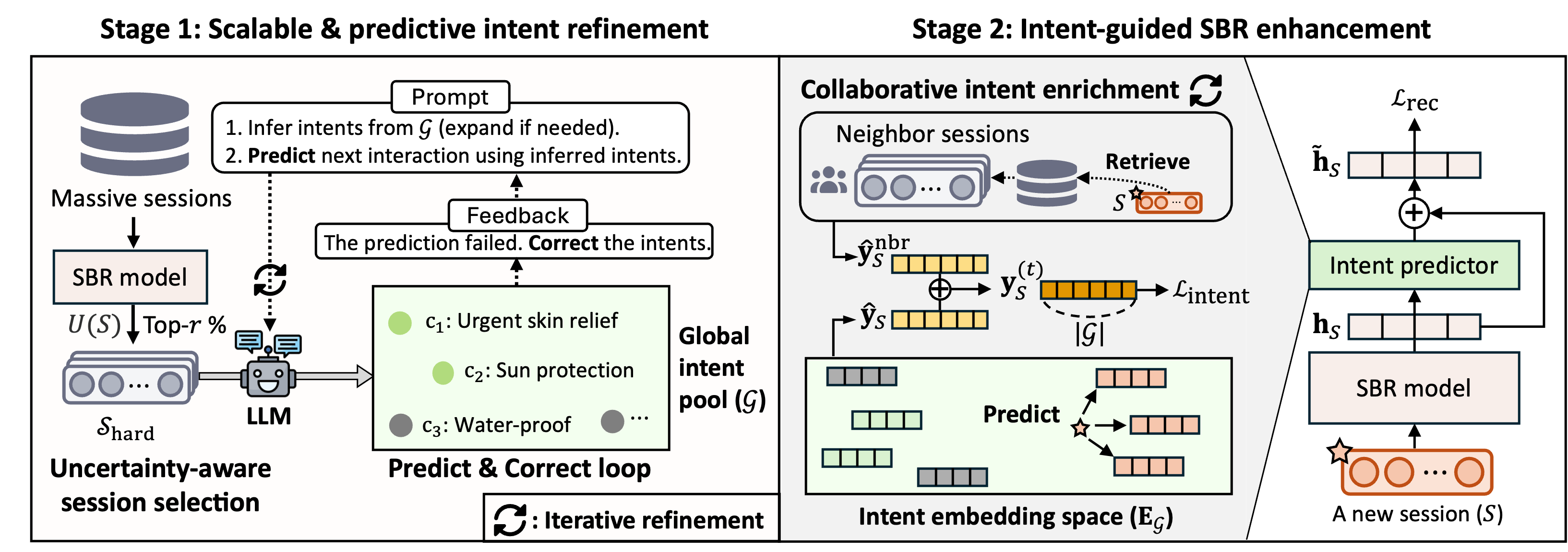}
    \caption{Overview of the \proposed framework. Best viewed in color.
    }
    \label{fig:method}
    \vspace{-0.3cm}
\end{figure*}

We propose \proposed, a two-stage framework that enhances the SBR model with LLM-generated intents.
In Stage 1, we generate session-level multiple intents by harnessing LLMs (\cref{sub:stage1}).
In Stage 2, we integrate these intents into the SBR model to enhance recommendation performance (\cref{sub:stage2}).
Finally, we describe the optimization process (\cref{sub:objective_function}).
Figure~\ref{fig:method} shows an overview of \proposed.

\subsection{Stage 1: Scalable and Predictive Intent Refinement via LLMs}\label{sub:stage1}
The goal of this stage is to infer a set of session-level intents that provide informative signals for predicting user behavior in a scalable manner, which will be used to enhance SBR models.
We first identify training sessions with high uncertainty using a conventional SBR model to selectively invoke LLM reasoning (\cref{subsub:uncertainty_select}).
Then, we construct a global intent pool to serve as a shared and constrained space of intents, and apply an iterative predict-and-correct loop to obtain reliable and informative intents (\cref{subsub:session_level}).


\subsubsection{\textbf{Uncertainty-aware hard session selection}}
\label{subsub:uncertainty_select}
To reduce the substantial computational cost of generating LLM-based intents for all sessions \textbf{(C2)}, we selectively allocate the LLM’s reasoning to the hard sessions via an uncertainty-aware selection strategy.
This approach employs a base SBR model, warmed up exclusively on the training sessions, to identify sessions with high uncertainty.
We prioritize these sessions—where underlying patterns are difficult to discern from raw interactions alone—as they stand to benefit most from the nuanced reasoning of LLMs.

For each training session $S$, we quantify its uncertainty via two metrics: session entropy $H_{\text{sess}}(S)$ and prediction difficulty $H_{\text{pred}}(S)$.
\begin{equation}\label{eq:entropy}
    \begin{aligned}
        H_{\text{sess}}(S) &= -\sum_{i \in S} P_{\theta}(i\;|\;S) \log P_{\theta}(i\;|\;S), \\ 
        H_{\text{pred}}(S) &= -\log P_{\theta}({i_N}\;|\;S_{1:N-1}), \\
    \end{aligned}
\end{equation}
\noindent where $P_\theta(i | S) = \frac{\exp(\mathbf{h}_S^\top \mathbf{e}_i)}{\sum_{j \in \mathcal{I}} \exp(\mathbf{h}_S^\top \mathbf{e}_j)}$, 
and $i_N$ is the held-out last item of the session $S$.
$\mathbf{h}_S \in \mathbb{R}^d$ is the session representation\footnote{\proposed seamlessly integrates with various conventional SBR models that produce session representations.
For instance, $\mathbf{h}_S $ corresponds to the hidden state of the last item in the session for both RNN-based~\cite{gru4rec,narm} and transformer-based models~\cite{sasrec,bert4rec}.} and $\mathbf{e}_i \in \mathbb{R}^d$ is the embedding of item $i$.
Specifically, $H_{\text{sess}}(S)$ reflects internal interest inconsistency, where high values indicate diverse item interactions within a session.
In contrast, $H_{\text{pred}}(S)$ measures the model’s lack of confidence in predicting $i_N$.

Since $H_{\text{sess}}(S)$ and $H_{\text{pred}}(S)$ capture complementary aspects of session uncertainty, we combine them into a unified score $U(S)$ using reciprocal rank fusion~\cite{cormack2009reciprocal, kweon2025pairsem} for scale-invariant integration.
\begin{equation}
    U(S) = \frac{1}{k + r_\text{sess}(S)} + \frac{1}{k + r_\text{pred}(S)}.
\end{equation}
$r_\text{sess}(S)$ and $r_\text{pred}(S)$ denote the ranks of session $S$ among all training sessions sorted in descending order of $H_{\text{sess}}(S)$ and $H_{\text{pred}}(S)$, respectively.
We set the damping constant $k=60$ following \cite{bruch2023analysis}.

Using $U(S)$, we select the top-$r\%$ most ambiguous sessions, denoted as $\mathcal{S}_{\text{hard}}$.
In this work, we set $r=10$. 
Then, the LLM infers intents exclusively for $\mathcal{S}_{\text{hard}}$, thereby substantially improving the scalability of our framework.
We show that our strategy achieves significantly reduced training time and robust performance across different values of $r$.
We provide a detailed analysis in \cref{sub:efficiency_analysis}.




\subsubsection{\textbf{Predictive intent refinement}}
\label{subsub:session_level}
A straightforward way to acquire intents is to simply instruct LLMs to infer them for each session \cite{lian2025egrec, LLM4SBR, PO4ISR}.
However, due to session context scarcity \textbf{(C1)}, this approach has several limitations.
First, it lacks control over the generation; there is always a risk of hallucination, and LLMs often use different terms to refer to the same concept (e.g., “eco-friendly” vs “environmentally friendly” skincare), making it difficult to handle the outputs consistently.
Second, even when a generated intent appears reasonable, it often remains overly broad (e.g., “summer necessities”), providing limited insight into user behavior.

As a solution, we propose a novel approach to improve the reliability and informativeness of LLM-generated intents.
We first construct the \textit{Global Intent Pool}, a shared and constrained set of intents.
We then use LLMs to pinpoint the most relevant intents from the pool and expand it only when no appropriate candidates are available.
By grounding intent generation in the intent pool, we constrain the output space of LLMs, reducing hallucination and improving consistency in expression.
Furthermore, the generated intents are iteratively refined and validated through the \textit{Predict-and-Correct Loop} to ensure the quality of the intents.


\smallsection{Global Intent Pool (GIP).}
We introduce GIP, denoted as $\mathcal{G}$, to constrain the LLM’s output space for session-level intent identification.
We first initialize $\mathcal{G}$ by instructing the LLM using a prompt $\mathcal{P}_\text{GIP}$ conditioned on the recommendation domain $D$ (e.g., Beauty): ``\texttt{Generate a minimal set of distinct intents from general to domain-specific for <$D$>.}'' This process yields basic, domain-relevant concepts (e.g., ``anti-aging'', ``fragrance-free products''), formally defined as:
\begin{equation}
    \mathcal{G} = \mathrm{LLM}(\mathcal{P}_\text{GIP}(D)).
\end{equation}
After initialization, the P\&C loop is executed per session $S$: The LLM identifies relevant intents from $\mathcal{G}$ or generates a new intent $c'$ only when no suitable candidates exist. 
$c'$ is integrated into $\mathcal{G}$ only if it contributes to an accurate prediction of user behavior via 
$\mathcal{G} \leftarrow \mathcal{G} \cup \{c'\}$.
By sharing predefined intents and generating new ones only as needed, GIP constrains the proliferation of redundant concepts, maintaining a compact and discriminative intent set.\footnote{In our experiments, the final number of generated intents is 120, 135, and 118 for Beauty, Yelp, and Book, respectively.}

\smallsection{Predict‑and‑Correct (P\&C) loop.}
We introduce the P\&C loop, which iteratively refines and validates inferred intents to ensure high informativeness.
We define intent informativeness through recommendation utility: an intent is valid only if it facilitates the accurate prediction of subsequent user behaviors.

For each session $S \in \mathcal{S}_{\text{hard}}$, we first construct a small \textit{intent validation task} by holding out the last item $i_N$ from $S$. 
We then define a candidate set $I_S = \{i_{N}, i_{\text{out}_1}, \dots, i_{\text{out}_M}\}$, 
where $\{i_{\text{out}_m}\}_{m=1}^M$ are $M$ out-of-session items not observed in $S$.\footnote{As the simplest choice, we randomly select the out-of-session items. $M$ is set to $5$.}
The LLM is instructed to infer underlying intents from session $S$, and then \textbf{predict} the item $\hat{i} \in I_S$ that is most likely to be the held-out item based on the inferred intents.
If $\hat{i}$ matches $i_N$, the inferred intents are accepted; otherwise, they are rejected.
The LLM is then provided with verbal feedback to rectify its reasoning process and \textbf{correct} the intents.

Specifically, the initial iteration begins by obtaining the feature matrix for the candidate set, $\mathbf{X}_{I_S} = \{\mathbf{x}_{N}, \mathbf{x}_{\text{out}_1}, \dots, \mathbf{x}_{\text{out}_M}\}$, to construct the starting input $V^{(0)}_S = (\mathbf{X}_S, \mathbf{X}_{I_S}, \mathcal{G})$. The LLM then jointly generates a set of session intents $\mathcal{C}_S^{(0)}$ and predicts an item $\hat{i}$ via:
%
%
%
\begin{equation}\label{eq:PC}
    \mathcal{C}^{(0)}_S, \hat{i} \gets \mathrm{LLM}(\mathcal{P}_{P\&C}(V_S^{(0)})), 
\end{equation}
where the two tasks in prompt $\mathcal{P}_{P\&C}$ are defined as follows.
\begin{tcolorbox}[colback=gray!10, colframe=black, boxrule=0.5pt, arc=0pt, left=2pt, right=2pt, top=2pt, bottom=2pt]
$\mathcal{P}_{P\&C}$ = ``\texttt{(1) Infer multiple session-level intents $\mathcal{C}^{(0)}_S$ from $\mathcal{G}$; generate novel and distinct intents only when no suitable candidate is identified. 
(2) Predict the held-out item from $I_S$ using $\mathcal{C}^{(0)}_S$.}''
\end{tcolorbox}
If $\hat{i} = i_{N}$, we terminate the loop and accept $\mathcal{C}^{(0)}_S$ as the final predictive intent set for session $S$, denoted as $\mathcal{C}^{\scriptscriptstyle\mathrm{LLM}}_S$. 
Any newly generated intents in $\mathcal{C}^{\scriptscriptstyle\mathrm{LLM}}_S$ are added~to~$\mathcal{G}$. 
Otherwise, we append explicit verbal feedback $R^{(0)}_S$ (e.g., ``\texttt{The prediction $\hat i$ failed. Correct the intents $\mathcal{C}^{(0)}_S$ and retry.}") to the previous input:
\begin{equation}\label{eq:P1}
    \begin{aligned}
        V^{(1)}_S = (V^{(0)}_S, R^{(0)}_S).
    \end{aligned}
\end{equation}
This loop continues up to $T$ iterations, and any sessions that remain inconclusive within $T$ iterations are subsequently handled during training (\cref{subsub:int_enrich}).
In this work, we set $T=3$.
The number of intents and iterations required for successful prediction varies across sessions and is automatically determined by the LLM based on the inherent complexity of each user’s behavior.
The overall procedure of the P\&C loop is provided in Algorithm~\ref{al_P&C}.

\input{algorithms/al_stage1}
\vspace{0.35cm}

\subsection{Stage 2: Intent-guided SBR Enhancement}\label{sub:stage2}
This stage aims to enhance SBR models using LLM-generated intents.
A naive approach is to treat them as static input features;
for example, one may obtain their textual embeddings and feed them into the model.
However, this is suboptimal for two reasons: 
(1) it is inapplicable to sessions without pre-generated intents, necessitating costly LLM inference at test time;
and (2) it fails to reflect the varying importance of multiple intents within a session.

We first introduce a multi-intent prediction to identify diverse user motivation (\cref{subsub:mi_pred_fusion}).
We then present collaborative intent enrichment to effectively impute missing intent signals (\cref{subsub:int_enrich}).
Finally, these multi-faceted intents are integrated into the SBR model through an intent fusion mechanism (\cref{subsub:fuse}).

\subsubsection{\textbf{Intent predictor}}
\label{subsub:mi_pred_fusion}
While intents are generated for hard sessions in $\mathcal{S}_{\text{hard}}$ during Stage 1, the discovered intent concepts in $\mathcal{G}$ can also be used to better understand other sessions.
To avoid repetitive LLM calls, we train a lightweight intent predictor \cite{kang2024improving,kweon2025topic} to generalize the LLM’s reasoning from $\mathcal{S}_{\text{hard}}$ to remaining sessions in $\mathcal{S}_\text{train} \setminus \mathcal{S}_\text{hard}$.
This enables scalable intent acquisition across the entire dataset and also avoids LLM dependency at test time \textbf{(C2)}.

Let $\mathbf{E}_\mathcal{G} = \{\mathbf{e}_c\}_{c \in \mathcal{G}}$ be the set of intent embeddings, where $\mathbf{e}_c \in \mathbb{R}^d$ is the embedding of intent $c$.
To model the session-intent relevance, we employ the query-key-value design~\cite{transformer}, where the relevance is computed by matching a query and a key.
Specifically, the session query $\mathbf{q}_{\scriptscriptstyle S}$, the intent key $\mathbf{k}_c$, and the value $\mathbf{v}_c$ are obtained as follows:
\begin{equation}\label{eq:qkv}
    \begin{aligned}
        \mathbf{q}_{\scriptscriptstyle S} = \mathbf{W}_{q}\mathbf{h}_{\scriptscriptstyle S} \in \mathbb{R}^{d}, \quad
        \mathbf{k}_c = \mathbf{W}_{k}\mathbf{e}_c \in \mathbb{R}^{d}, \quad
        \mathbf{v}_c = \mathbf{W}_{v}\mathbf{e}_c \in \mathbb{R}^{d},
    \end{aligned}
\end{equation}
where $\mathbf{W}_{q}, \mathbf{W}_{k}, \mathbf{W}_{v} \in \mathbb{R}^{d \times d}$ are the projection matrices for queries, keys, and values, respectively.
We compute the session-intent relevance, denoted as $\mathbf{\hat{y}}_{\scriptscriptstyle S} = [\hat{y}_{\scriptscriptstyle Sc}]_{c \in \mathcal{G}}$, as follows: 
\begin{equation}\label{eq:y_hat}
    \begin{aligned}
        \hat{y}_{\scriptscriptstyle Sc} = \sigma(\mathbf{q}_{\scriptscriptstyle S}^\top\mathbf{k}_c) \in [0,1],
    \end{aligned}
\end{equation}
where $\sigma$ denotes the sigmoid function~\cite{sigmoid}.
Unlike the softmax function~\cite{softmax}, which enforces the mutual exclusivity of intents via $\sum_c\hat{y}_{\scriptscriptstyle Sc}=1$, the sigmoid activation provides more flexibility to capture multiple relevant intents non-exclusively.
The session-intent relevance is then learned by the following~loss:
\begin{equation}
\begin{aligned}\label{eq:intent_loss}
        \mathcal{L}_{\text{intent}} = -\frac{1}{|\mathcal{G}|}
        \sum_{c \in \mathcal{G}} \Big[y^{(t)}_{\scriptscriptstyle Sc}\log\hat{y}_{\scriptscriptstyle Sc} + (1-y^{(t)}_{\scriptscriptstyle Sc})\log(1-\hat{y}_{\scriptscriptstyle Sc})\Big],
\end{aligned}
\end{equation}
where $\mathbf{y}^{(t)}_{\scriptscriptstyle S} = [y^{(t)}_{\scriptscriptstyle Sc}]_{c \in \mathcal{G}}$ denotes the intent labels at training epoch~$t$.
In the early stages of training, the labels are derived from LLM-generated intents $\{\mathcal{C}^{\scriptscriptstyle\mathrm{LLM}}_S\}_{S \in \mathcal{S}_\text{hard}}$.
As training progresses, we incorporate enriched intent labels to cover all training sessions, a process further elaborated in the following subsection.

\subsubsection{\textbf{Collaborative intent enrichment}}\label{subsub:int_enrich}
To accurately train the intent predictor, we propose a collaborative intent enrichment strategy.
This approach iteratively refines intent supervision by leveraging neighboring sessions with similar behavioral patterns.

At the beginning of training, we assign initial intent labels $\mathbf{y}^{(0)}_{\scriptscriptstyle S} =[y^{(0)}_{\scriptscriptstyle Sc}]_{c \in \mathcal{G}}$ for sessions where LLM-derived intents are available. 
\begin{equation}
    \begin{aligned}
        y^{(0)}_{\scriptscriptstyle Sc} =
        \begin{cases}
        1, & c \in \mathcal{C}^{\scriptscriptstyle\mathrm{LLM}}_S,\\
        0, & \text{otherwise}.
        \end{cases}
    \end{aligned}
\end{equation}
During training, we progressively enrich missing intent labels using predictions from the intent predictor (i.e., $\mathbf{\hat{y}_{\scriptscriptstyle S}}$ in Eq.~\eqref{eq:y_hat}).
This procedure is inspired by self-training \cite{selftrain1, selftrain2}, a well-established semi-supervised learning paradigm where a model learns from its own predictions on unlabeled data.

However, predictions made for individual sessions can be unstable and noisy, particularly in the early stages of training.
To improve robustness, we propagate intent information from behaviorally similar sessions.
Specifically, given a session representation $\mathbf{h}{\scriptscriptstyle S}$, we first retrieve its top-$K$ most similar sessions within the same mini-batch $\mathcal{B}$ based on cosine similarity:
\begin{equation}\label{eq:neighbor}
    \begin{aligned}
        \mathcal{N}(S) = \mathrm{TopK}(\{\text{sim}_{\scriptscriptstyle S'}\}_{S' \in \mathcal{B}, S' \neq S}),
        \quad
        \text{sim}_{\scriptscriptstyle S'} = \mathrm{cos}(\mathbf{h}_{\scriptscriptstyle S}, \mathbf{h}_{\scriptscriptstyle S'}).
    \end{aligned}
\end{equation}
We then aggregate neighbor intent predictions into $\hat{\mathbf{y}}^{\text{nbr}}_S$ using similarity weights $w_{\scriptscriptstyle S'}$\footnote{$w_{\scriptscriptstyle S'} = {\mathrm{exp}(\text{sim}_{\scriptscriptstyle S'})}/{\sum_{S''\in\mathcal{N}(S)}\mathrm{exp}(\text{sim}_{\scriptscriptstyle S''})}$}:

\begin{equation}
    \begin{aligned}
    \hat{\mathbf{y}}^{\text{nbr}}_{\scriptscriptstyle S} \!= \sum_{S'\in\mathcal{N}(S)}\!w_{S'} \cdot\hat{\mathbf{y}}_{S'}.
    \end{aligned}
\end{equation}
At training epoch $t$, we obtain the enriched intent labels $\mathbf{y}^{(t)}_{\scriptscriptstyle S}$ by averaging the session-level and neighbor-aggregated predictions:
\begin{equation}\label{eq:intent_update}
    \begin{aligned}
    \mathbf{y}^{(t)}_{\scriptscriptstyle S} =  \frac{1}{2}\bigl(\hat{\mathbf{y}}_{\scriptscriptstyle S} + \hat{\mathbf{y}}^{\text{nbr}}_{\scriptscriptstyle S}).
    \end{aligned}
\end{equation}
These enriched labels $\mathbf{y}^{(t)}_{\scriptscriptstyle S}$ are then used as targets for $\mathcal{L}_{\text{intent}}$ (Eq.~\eqref{eq:intent_loss}).

The enriched labels $\mathbf{y}^{(t)}_{\scriptscriptstyle S}$ provide richer supervision in two aspects.
First, when explicit intent supervision is missing, they supply pseudo-intents inferred from behaviorally similar sessions.
Notably, compared to relying solely on a single session-level prediction $\hat{\mathbf{y}}_{\scriptscriptstyle S}$, our neighborhood-based enrichment yields a more stable and reliable signal.
Second, unlike the initial binary labels $\mathbf{y}^{(0)}_{\scriptscriptstyle S}$, the enriched labels $\mathbf{y}^{(t)}_{\scriptscriptstyle S}$ encode soft probabilities that reflect the importance of each intent for a given session, providing more nuanced supervision for learning.
As updating the labels every epoch is both inefficient and unnecessary, we update $\mathbf{y}^{(t)}_{\scriptscriptstyle S}$ every $\rho$ epochs. We set $\rho$ to 5.

\subsubsection{\textbf{Fusing multi-intent information.}}\label{subsub:fuse}
We enhance the session representation with intents based on their varying importance to the session. 
We create the guidance representation $\mathbf{g}_{\scriptscriptstyle S}$ as follows:
\begin{equation}\label{eq:g}
    \begin{aligned}
        \mathbf{g}_{\scriptscriptstyle S} = \mathrm{Dropout}\Big( \mathrm{LayerNorm}\big(\mathbf{h}_{\scriptscriptstyle S} + \sum_{c \in \mathcal{G}}\phi(\hat{y}_{\scriptscriptstyle Sc})\mathbf{v}_c\big)\Big),
    \end{aligned}
\end{equation}
where $\phi(a) = a \cdot\mathbb{I}[a > \tau]$ is a filtering function with threshold $\tau$.
$\mathbf{v}_c$ is the intent-aspect value vector defined in Eq~\eqref{eq:qkv}.
This filtering step allows the model to focus on intents above a certain degree of relevance, while preserving their relative importance.
In this work, we set $\tau = 0.5$.
The final intent-enhanced representation is obtained as:
$\Tilde{\mathbf{h}}_{\scriptscriptstyle S} = \mathbf{h}_{\scriptscriptstyle S} +  \mathbf{g}_{\scriptscriptstyle S}$.
The standard next-item prediction loss is adopted:
\begin{equation}\label{eq:rec_loss}
\hspace{-0.1cm} 
    \begin{aligned}
        \mathcal{L}_{\text{rec}} = -\sum_{S \in \mathcal{S}_\text{train}}\Big[ \log \sigma  (\mathbf{e}_{i_{\text{pos}}}^\top\Tilde{\mathbf{h}}_S )+\!\;\sum_{i_\text{neg}}
        \log \big(1-\sigma  (\mathbf{e}^\top_{i_{\text{neg}}}\Tilde{\mathbf{h}}_S )\big)
        \Big].
    \end{aligned}
\end{equation}
For each training session, we sample two negative items: one randomly sampled negative and one hard negative selected from items mispredicted in the P\&C loop.
If no hard negatives are available, both negatives are sampled at random \cite{kweon2024doubly}.



%

\subsection{Optimization of \proposed}
\label{sub:objective_function}
The final learning objective of \proposed is defined as:
\begin{equation}\label{eq:optimize}
    \begin{aligned}
        \min_{\substack{\theta}} \mathcal{L}_\text{Rec} + \lambda_{\text{intent}} \mathcal{L}_\text{intent}  + \lambda_{\text{ortho}} \mathcal{L}_\text{ortho},
    \end{aligned}
\end{equation}
\noindent where $\theta = \{\mathrm{SBR}(\cdot), \mathbf{W}_q, \mathbf{W}_k, \mathbf{W}_v, \mathbf{E}_\mathcal{G}\}$ denotes the parameters of \proposed.
We additionally adopt an intent-orthogonality loss, which has been widely used in prior intent-aware methods~\cite{yang2023based, ELCRec}, to encourage different intent embeddings to capture distinct semantic aspects:
$\mathcal{L}_{\text{ortho}} = -\frac{1}{|\mathcal{G}|(|\mathcal{G}|-1)}\sum^{|\mathcal{G}|}_{i=1} \sum^{|\mathcal{G}|}_{j=1, j \neq i} ||\hat{\mathbf{e}}_{c_i} - \hat{\mathbf{e}}_{c_j}||^2_2$, where $\hat{\mathbf{e}}_c$ denotes the L2-normalized embedding of intent $c$.


\smallsection{Inference.}
At test time, \proposed does not invoke LLM to generate intents.
Instead, \proposed leverages the trained intent predictor to identify relevant intents and incorporate them into the prediction process, without incurring additional LLM latency for intent extraction.
The next item is predicted as: $\hat{i} = \operatorname{argmax}_{j \in \mathcal{I}}\, \sigma(\mathbf{e}^\top_{j}\;        
\Tilde{\mathbf{h}}_{\scriptscriptstyle S})$.

%% file: algorithms/al_stage1.tex
\begin{algorithm}[t]
\footnotesize
\DontPrintSemicolon
\SetKwInOut{Input}{Input}
\SetKwInOut{Output}{Output}

\Input{Hard session $S \in \mathcal{S}_\text{hard}$, GIP $\mathcal{G}$, max iterations $T$}
\Output{Final predictive intent set $\mathcal{C}^{\scriptscriptstyle\mathrm{LLM}}_S$, Updated GIP $\mathcal{G}$}
\BlankLine
Construct the candidate item set $I_S$ \\
Construct the initial LLM input $ V^{(0)}_S \leftarrow (\mathbf{X}_S, \mathbf{X}_{I_S}, \mathcal{G})$ \\
Initialize the set of intents $\mathcal{C}^{\scriptscriptstyle\mathrm{LLM}}_S \leftarrow\emptyset$ \\

\SetInd{0.2em}{0.8em}
\For{$t = 0$ \KwTo $T\!-1\!$}{
    \tcp*[h]{Infer intents and predict the held-out item} \\
    $\mathcal{C}^{(t)}_S, \hat{i} \gets \mathrm{LLM}(\mathcal{P}_{P\&C}(V_S^{(t)}))$ \Comment{Eq. \eqref{eq:PC}} \\

    \eIf{$\hat{i} = i_N$}{
Accept the final intent set $\mathcal{C}^{\scriptscriptstyle\mathrm{LLM}}_S \gets \mathcal{C}^{(t)}_S$ \\
\textbf{break}
}{

$R^{(t)}_S \gets$ (${\footnotesize \texttt{"The prediction } \hat{i} \texttt{ failed. Correct } \mathcal{C}^{(t)}_S \texttt{ and retry."}}$) \\
Update the LLM input $V^{(t+1)}_S \gets (V^{(t)}_S, R^{(t)}_S)$ \Comment{Eq.~\eqref{eq:P1}}
}
}
\tcp*[h]{Update GIP with newly generated intents} \\
\vspace{0.1cm}
$\mathcal{G} \leftarrow \mathcal{G} \cup \mathcal{C}^{\scriptscriptstyle\mathrm{LLM}}_S$ \\
\vspace{0.1cm}
\Return $\mathcal{C}^{\scriptscriptstyle\mathrm{LLM}}_S, \mathcal{G}$
\caption{P\&C loop}
\label{al_P&C}
\end{algorithm}
\vspace{-0.4cm}

%% file: sections/050Experiment_setup.tex






\subsubsection{\textbf{Datasets.}} 
We use three real‑world datasets: Beauty (Amazon)~\cite{mcauley2015image}, Yelp~\cite{yelp_open_dataset}, and Books (Amazon)~\cite{mcauley2015image}, following \cite{LLM4SBR, ELCRec,FAPAT,RLMRec,LLMESR}.
For preprocessing, we follow the conventional protocol~\cite{LLM4SBR, MiasRec, FAPAT} and apply $5$-core filtering.
For data splitting, we sort all sessions by the timestamp of their last event and use the 80th and 90th percentiles of these timestamps as cut‑off points to split the data into training, validation, and test sets in an 8:1:1 ratio.
Table \ref{tab:dataset} presents detailed statistics. 

\input{tables/tab_dataset}
\input{tables/tab_maintable_fix2} 

\vspace{-0.1cm}
\subsubsection{\textbf{Baselines}}
We group the baselines into three categories and compare \proposed with representative methods from each category.
\begin{enumerate}[leftmargin=*]\vspace{-\topsep}
    \item \textbf{Implicit modeling}: infers intents in an unsupervised~manner.
    \begin{itemize}[leftmargin=*]
        \item \textbf{ELCRec}~\cite{ELCRec} learns intent embeddings via a push–pull loss to enrich representations with the dominant intent.
        \item \textbf{MiaSRec}~\cite{MiasRec} dynamically models and selects multiple intents per session to guide item distribution.
    \end{itemize}
    
    \item \textbf{Explicit modeling}: infers explicit intents from item metadata.
    \begin{itemize}[leftmargin=*]
        \item \textbf{FAPAT}~\cite{FAPAT} models intents from frequent attribute patterns and then uses them to augment session representations.
        \item \textbf{C\textsc{a}F\textsc{e}}~\cite{CAFE} treats item categories as explicit intents and captures both coarse- and fine-grain sequential dynamics.
    \end{itemize}

    \item \textbf{LLM-enhanced CRM}: uses outputs from LLMs (e.g., intents, text embeddings or profiles) to enhance CRMs.
    \begin{itemize}[leftmargin=*]
        \item \textbf{LLM4SBR}~\cite{LLM4SBR} uses an LLM to infer long‑ and short‑term view intents from item titles within each session and incorporates them into the SBR model.
        \item \textbf{LLM-ESR}~\cite{LLMESR} converts item‑level features into LLM semantic embeddings and fuses this semantic view with collaborative view from the CRM in a dual‑view framework.
        \item \textbf{RLMRec}~\cite{RLMRec} leverages LLM‑derived semantic representations of user/item profiles to enhance CRM learning via two variants: \textbf{RLMRec‑Con}  projects semantic features into the CRM space, and \textbf{RLMRec‑Gen} does the reverse.
        \item \textbf{ProfileRec} denotes the method proposed in~\cite{ProfileRec}. It is a state-of-the-art LLM-based profiling method that employs a self-reflection process to generate reliable dual profiles (self and neighbor) for each user and item.
        These profiles are aligned via contrastive learning to enhance CRMs.
    \end{itemize}
\end{enumerate}\vspace{-\topsep}
Note that we do not include methods that use LLMs as standalone CRMs~\cite{PO4ISR, wang2025re2llm} in our baselines, as it is a different research direction. 
Instead, we provide a separate analysis in Table~\ref{tab:inference_efficiency}.
We use \texttt{GPT‑3.5‑turbo} for all LLM-based methods.
Results obtained with \texttt{Llama-3.3-70B-Instruct} are provided in Table~\ref{tab:llama}.
For backbone SBR models, we use \textbf{SASRec}~\cite{sasrec} and \textbf{BERT4Rec}~\cite{bert4rec} as many baselines commonly employ Transformer~\cite{transformer} architectures.



\subsubsection{\textbf{Model variants.}}
We note that all LLM‑based baselines (i.e., LLM4SBR, LLM-ESR, RLMRec, and ProfileRec) incorporate LLM embeddings (LE) of item metadata as additional item features.
\proposed is compatible with this design. 
To ensure a fair comparison, we also report the results of this variant, denoted as \textbf{\proposed\!+LE}.
Specifically, for each item in a session, we obtain a trainable base embedding and a frozen LLM-derived embedding.
For a fair comparison under a fixed dimensionality budget, the LLM-derived embedding is projected to $d/2$ dimensions using a two-layer MLP, and combined with the base embedding of the same dimensionality to form a $d$-dimensional session representation.
Each branch is encoded using the same SBR model, and the resulting outputs are concatenated to obtain the final session embedding.
For LLM embeddings, we employ OpenAI's \texttt{text-embedding-ada-002}, except in Table~\ref{tab:llama} where embeddings from the corresponding Llama model are adopted.


\vspace{-0.1cm}
\subsubsection{\textbf{Evaluation metrics.}} 
We evaluate recommendation quality using Hit Rate (H)~\cite{Hit} and NDCG (N)~\cite{NDCG} at cutoffs \{5, 10, 20\}, following prior studies~\cite{ELCRec, PO4ISR, RLMRec, FAPAT, LLMESR}.
All results are averaged over five independent runs with different random seeds.


\subsubsection{\textbf{Implementation details.}} 
We implement all models in PyTorch with CUDA on RTX 3090 GPUs and AMD EPYC 7413 CPUs. Hyperparameters are tuned via grid search on the validation set. The learning rate is chosen from $\{1\text{e-}3, 2\text{e-}3, 4\text{e-}3\}$, $L_2$ regularization from $\{1\text{e-}5, 1\text{e-}4, 1\text{e-}3, 1\text{e-}2\}$, and the dropout ratio from $\{0.1, 0.2, 0.3, \\ 0.4\}$. Transformer backbones use 2 layers, 2 heads, and an embedding dimension of 64. For \proposed, the number of top-$K$ neighbor sessions is chosen from $\{5, 10, 15, 20, 25\}$, and $\lambda_{\text{intent}}$, $\lambda_{\text{ortho}}$ are chosen from $\{1\text{e-}4, 1\text{e-}3, 1\text{e-}2, 1\text{e-}1\}$. Baseline hyperparameter ranges follow the original papers.

%% file: tables/tab_dataset.tex

\begin{table}[t]
\caption{Dataset statistics after preprocessing.}
\centering
\renewcommand{\arraystretch}{0.6}
\resizebox{\columnwidth}{!}{%
\begin{tabular}{cc|ccccc}
\toprule
\multicolumn{2}{c|}{\textbf{Dataset}} & \textbf{\#Sessions} & \textbf{\#Items} & \textbf{\begin{tabular}[c]{@{}c@{}} \textbf{\#Inter-} \\ \textbf{actions}\end{tabular}} & \textbf{\begin{tabular}[c]{@{}c@{}} \textbf{Avg.} \\ \textbf{Length}\end{tabular}} & \textbf{Metadata} \\ \midrule\midrule
\multicolumn{1}{c|}{\multirow{-1.4}{*}{\rotatebox{90}{\textbf{Beauty}}}}
 & Total & 7,524 & 2,351 & 31,538 & 4.19 & \multirow{4}{*}{\begin{tabular}[c]{@{}c@{}}description,\\ categories, rating,\\ title, review, price\end{tabular}} \\
\multicolumn{1}{c|}{} & Train & 6,448 & 2,351 & 27,130 & 4.21 &  \\
\multicolumn{1}{c|}{} & Valid & 611   & 1,126 & 2,480  & 4.10 &  \\
\multicolumn{1}{c|}{} & Test  & 465   & 1,085 & 1,928  & 4.15 &  \\ \midrule
\multicolumn{1}{c|}{\multirow{5.5}{*}{\rotatebox{90}{\textbf{Yelp}}}}
 & Total & 41,892 & 8,701 & 152,818 & 3.65 & \multirow{4}{*}{\begin{tabular}[c]{@{}c@{}}name,\\ stars,\\ categories, \\ text\end{tabular}} \\
\multicolumn{1}{c|}{} & Train & 34,296 & 8,701 & 125,181 & 3.65 &  \\
\multicolumn{1}{c|}{} & Valid & 3,985  & 5,112 & 14,453  & 3.63 &  \\
\multicolumn{1}{c|}{} & Test  & 3,611  & 4,621 & 13,184  & 3.65 &  \\ \midrule
\multicolumn{1}{c|}{\multirow{5.5}{*}{\rotatebox{90}{\textbf{Book}}}}
 & Total & 20,102 & 9,578 & 141,510 & 7.03 & \multirow{4}{*}{\begin{tabular}[c]{@{}c@{}}description,\\ categories, rating, \\ title, review, price\end{tabular}} \\
\multicolumn{1}{c|}{} & Train & 16,864 & 9,578 & 118,638 & 7.03 &  \\
\multicolumn{1}{c|}{} & Valid & 1,723  & 4,957 & 12,233  & 7.10 &  \\
\multicolumn{1}{c|}{} & Test  & 1,515  & 4,919 & 10,639  & 7.02 &  \\ \bottomrule
\end{tabular}
}
\label{tab:dataset}
\end{table}

%% file: tables/tab_maintable_fix2.tex
\begin{table*}[ht!] \vspace{-0.1cm}
\caption{Overall performance comparison. * indicates $p< 0.05$ for the paired t-test against the best baseline.}
\centering
\renewcommand{\arraystretch}{0.5}
\resizebox{0.93\linewidth}{!}{%
\begin{tabular}{cl|cccccc|cccccc} 
\toprule
\multicolumn{2}{c|}{}                                                                               & \multicolumn{6}{c|}{\textbf{SASRec}}                                                                                                                                                                                                                             & \multicolumn{6}{c}{\textbf{BERT4Rec}}                                                                                                                                                                                                                           \\
\multicolumn{2}{c|}{\multirow{-2}{*}{\textbf{Method}}}                                              & \textbf{H@5}                           & \textbf{N@5}                          & \textbf{H@10}                          & \textbf{N@10}                         & \textbf{H@20}                          & \textbf{N@20}                         & \textbf{H@5}                           & \textbf{N@5}                          & \textbf{H@10}                          & \textbf{N@10}                         & \textbf{H@20}                          & \textbf{N@20}                         \\ \midrule\midrule
\multicolumn{1}{c|}{}                                   & Backbone                         & 0.0198                                   & 0.0132                                   & 0.0297                                   & 0.0164                                   & 0.0456                                   & 0.0203                                   & 0.0194                                   & 0.0127                                   & 0.0310                                   & 0.0165                                   & 0.0538                                   & 0.0221                                   \\
\multicolumn{1}{c|}{}                                   & ELCRec                                    & {\ul 0.0267}                             & {\ul 0.0170}                             & 0.0357                                   & 0.0199                                   & 0.0590                                   & {\ul 0.0258}                             & 0.0228                                   & 0.0149                                   & 0.0417                                   & 0.0210                                   & 0.0650                                   & 0.0268                                   \\
\multicolumn{1}{c|}{}                                   & MiaSRec                                   & 0.0134                                   & 0.0073                                   & 0.0254                                   & 0.0111                                   & {\ul 0.0650}                                   & 0.0210                                   & 0.0151                                   & 0.0086                                   & 0.0301                                   & 0.0135                                   & 0.0714                                   & 0.0238                                   \\
\multicolumn{1}{c|}{}                                   & FAPAT                                     & 0.0250                                   & 0.0159                                   & {\ul 0.0383}                             & {\ul 0.0201}                             & 0.0594                                   & 0.0254                                   & 0.0241                                   & 0.0149                                   & 0.0379                                   & 0.0194                                   & 0.0628                                   & 0.0257                                   \\
\multicolumn{1}{c|}{}                                   & C\textsc{a}F\textsc{e}                                      & 0.0177                                   & 0.0108                                   & 0.0318                                   & 0.0154                                   & 0.0521                                   & 0.0204                                   & 0.0181                                   & 0.0112                                   & 0.0318                                   & 0.0156                                   & 0.0589                                   & 0.0225                                   \\
\multicolumn{1}{c|}{}                                   & LLM4SBR                                   & 0.0194                                   & 0.0122                                   & 0.0353                                   & 0.0174                                   & 0.0486                                   & 0.0208                                   & 0.0181                                   & 0.0116                                   & 0.0319                                   & 0.0161                                   & 0.0465                                   & 0.0198                                   \\
\multicolumn{1}{c|}{}                                   & LLM-ESR                                   & 0.0168                                   & 0.0104                                   & 0.0323                                   & 0.0154                                   & 0.0503                                   & 0.0199                                   & 0.0284                                   & 0.0185                                   & {\ul 0.0473}                             & 0.0244                                   & {\ul 0.0766}                             & 0.0317                                   \\
\multicolumn{1}{c|}{}                                   & RLMRec-Con                                & 0.0215                                   & 0.0139                                   & 0.0349                                   & 0.0180                                   & 0.0624                             & 0.0249                                   & 0.0284                                   & {\ul 0.0207}                             & 0.0447                                   & {\ul 0.0258}                             & 0.0684                                   & {\ul 0.0318}                             \\
\multicolumn{1}{c|}{}                                   & RLMRec-Gen                                & 0.0207                                   & 0.0123                                   & 0.0357                                   & 0.0172                                   & 0.0576                                   & 0.0227                                   & {\ul 0.0297}                             & 0.0191                                   & 0.0413                                   & 0.0229                                   & 0.0697                                   & 0.0299                                   \\
\multicolumn{1}{c|}{}                                   & ProfileRec                                & 0.0224                                   & 0.0146                                   & 0.0357                                   & 0.0188                                   & 0.0598                                   & 0.0250                                   & 0.0263                             & 0.0179                                   & 0.0413                                   & 0.0227                                   & 0.0671                                   & 0.0291                                   \\
\multicolumn{1}{c|}{}& \cellcolor[HTML]{D9D9D9}SPRINT & \cellcolor[HTML]{D9D9D9}\textbf{0.0331*} & \cellcolor[HTML]{D9D9D9}\textbf{0.0201*} & \cellcolor[HTML]{D9D9D9}\textbf{0.0503*} & \cellcolor[HTML]{D9D9D9}\textbf{0.0256*} & \cellcolor[HTML]{D9D9D9}\textbf{0.0740*}           & \cellcolor[HTML]{D9D9D9}\textbf{0.0316*} & \cellcolor[HTML]{D9D9D9}0.0297           & \cellcolor[HTML]{D9D9D9}0.0211 & \cellcolor[HTML]{D9D9D9}0.0435           & \cellcolor[HTML]{D9D9D9}0.0255           & \cellcolor[HTML]{D9D9D9}0.0680           & \cellcolor[HTML]{D9D9D9}0.0316                                     \\ 
\multicolumn{1}{c|}{\multirow{-16}{*}{\textbf{Beauty}}} & \cellcolor[HTML]{D9D9D9}SPRINT+LE  & \cellcolor[HTML]{D9D9D9}0.0293           & \cellcolor[HTML]{D9D9D9}0.0192          & \cellcolor[HTML]{D9D9D9}0.0430           & \cellcolor[HTML]{D9D9D9}0.0236           & \cellcolor[HTML]{D9D9D9}0.0568  & \cellcolor[HTML]{D9D9D9}0.0270           & \cellcolor[HTML]{D9D9D9}\textbf{0.0349*} & \cellcolor[HTML]{D9D9D9}\textbf{0.0244*}           & \cellcolor[HTML]{D9D9D9}\textbf{0.0503*} & \cellcolor[HTML]{D9D9D9}\textbf{0.0293*} & \cellcolor[HTML]{D9D9D9}\textbf{0.0817*}  & \cellcolor[HTML]{D9D9D9}\textbf{0.0372*}           \\ \midrule
\multicolumn{1}{c|}{}                                   & Backbone                         & 0.0323                                   & 0.0201                                   & 0.0554                                   & 0.0275                                   & 0.0867                                   & 0.0354                                   & 0.0483                                   & 0.0304                                   & 0.0831                                   & 0.0416                                   & 0.1356                                   & 0.0548                                   \\
\multicolumn{1}{c|}{}                                   & ELCRec                                    & 0.0355                                   & 0.0227                                   & 0.0612                                   & 0.0310                                   & 0.0985                                   & 0.0403                                   & 0.0533                                   & 0.0336                                   & 0.0877                                   & 0.0446                                   & 0.1338                                   & 0.0562                                   \\
\multicolumn{1}{c|}{}                                   & MiaSRec                                   & 0.0192                                  & 0.0112                                   & 0.0441                                   & 0.0190                                   & 0.0913                                   & 0.0310                                   & 0.0502                                   & 0.0275                                   & 0.0894                                   & 0.0400                                   & 0.1474                                   & 0.0547                                   \\
\multicolumn{1}{c|}{}                                   & FAPAT                                     & 0.0328                                   & 0.0206                                   & 0.0545                                   & 0.0275                                   & 0.0861                                   & 0.0355                                   & 0.0558                                   & 0.0356                                   & 0.0940                                   & 0.0478                                   & 0.1494                                   & 0.0617                                   \\
\multicolumn{1}{c|}{}                                   & C\textsc{a}F\textsc{e}                                      & 0.0351                                   & 0.0226                                   & 0.0606                                   & 0.0308                                   & 0.0974                                   & 0.0400                                   & 0.0564                                   & 0.0360                                   & 0.0909                                   & 0.0472                                   & 0.1409                                   & 0.0598                                   \\
\multicolumn{1}{c|}{}                                   & LLM4SBR                                   & 0.0325                                   & 0.0205                                   & 0.0552                                   & 0.0278                                   & 0.0911                                   & 0.0368                                   & 0.0552                                   & 0.0349                                   & 0.0925                                   & 0.0469                                   & 0.1467                                   & 0.0605                                   \\
\multicolumn{1}{c|}{}                                   & LLM-ESR                                   & 0.0363                                   & 0.0251                                   & 0.0606                                   & 0.0329                                   & 0.1031                                   & 0.0434                                   & 0.0542                                   & 0.0271                                   & 0.0883                                   & 0.0447                                   & 0.1422                                   & 0.0583                                   \\
\multicolumn{1}{c|}{}                                   & RLMRec-Con                                & 0.0351                             & 0.0225                             & 0.0571                             & 0.0295                             & 0.0894                                   & 0.0377                             & {\ul 0.0593}                             & {\ul 0.0377}                             & 0.0955                                   & 0.0493                                   & 0.1524                                   & 0.0636                                   \\
\multicolumn{1}{c|}{}                                   & RLMRec-Gen                                & 0.0349                                   & 0.0224                                   & 0.0552                                   & 0.0289                                   & 0.0901                             & 0.0376                                   & 0.0585                                   & 0.0375                                   & {\ul 0.0967}                             & {\ul 0.0498}                             & 0.1531                             & 0.0640                             \\
\multicolumn{1}{c|}{}                                   & ProfileRec                                & {\ul 0.0451}                                   & {\ul 0.0290}                                   & {\ul 0.0765}                                   & {\ul 0.0390}                                   & {\ul 0.1193}                             & {\ul 0.0498}                                   & 0.0588                                   & 0.0376                                   & 0.0961                             & 0.0497                             & {\ul 0.1540}                             & {\ul 0.0642}                             \\
\multicolumn{1}{c|}{} & \cellcolor[HTML]{D9D9D9}SPRINT & \cellcolor[HTML]{D9D9D9}\textbf{0.0495*} & \cellcolor[HTML]{D9D9D9}\textbf{0.0319*} & \cellcolor[HTML]{D9D9D9}0.0793 & \cellcolor[HTML]{D9D9D9}\textbf{0.0414*} & \cellcolor[HTML]{D9D9D9}\textbf{0.1272*} & \cellcolor[HTML]{D9D9D9}\textbf{0.0535*} & \cellcolor[HTML]{D9D9D9}0.0600 & \cellcolor[HTML]{D9D9D9}0.0377  & \cellcolor[HTML]{D9D9D9}0.1001           & \cellcolor[HTML]{D9D9D9}0.0506           & \cellcolor[HTML]{D9D9D9}0.1550           & \cellcolor[HTML]{D9D9D9}0.0644                                            \\ 
\multicolumn{1}{c|}{\multirow{-16}{*}{\textbf{Yelp}}}   & \cellcolor[HTML]{D9D9D9}SPRINT+LE  & \cellcolor[HTML]{D9D9D9}0.0476           & \cellcolor[HTML]{D9D9D9}0.0298           & \cellcolor[HTML]{D9D9D9}\textbf{0.0798*}           & \cellcolor[HTML]{D9D9D9}0.0402           & \cellcolor[HTML]{D9D9D9}0.1263          & \cellcolor[HTML]{D9D9D9}0.0519           & \cellcolor[HTML]{D9D9D9}\textbf{0.0628*}           & \cellcolor[HTML]{D9D9D9}\textbf{0.0404*}           & \cellcolor[HTML]{D9D9D9}\textbf{0.1037*} & \cellcolor[HTML]{D9D9D9}\textbf{0.0535*} & \cellcolor[HTML]{D9D9D9}\textbf{0.1577*} & \cellcolor[HTML]{D9D9D9}\textbf{0.0671*}  \\ \midrule
\multicolumn{1}{c|}{}                                   & Backbone                         & 0.0165                                   & 0.0101                                   & 0.0269                                   & 0.0135                                   & 0.0466                                   & 0.0184                                   & 0.0511                                   & 0.0366                                   & 0.0662                                   & 0.0414                                   & 0.0873                                   & 0.0467                                   \\
\multicolumn{1}{c|}{}                                   & ELCRec                                    & 0.0226                                   & 0.0141                                   & 0.0367                                   & 0.0185                                   & 0.0578                                   & 0.0239                                   & 0.0560                                   & 0.0396                                   & 0.0751                                   & 0.0458                                   & 0.0949                                   & 0.0507                                   \\
\multicolumn{1}{c|}{}                                   & MiaSRec                                   & 0.0162                                   & 0.0096                                   & 0.0309                                   & 0.0143                                   & 0.0692                                   & 0.0239                                   & 0.0354                                   & 0.0225                                   & 0.0665                                   & 0.0325                                   & 0.1102                                   & 0.0435                                   \\
\multicolumn{1}{c|}{}                                   & FAPAT                                     & 0.0203                                   & 0.0129                                   & 0.0318                                   & 0.0166                                   & 0.0488                                   & 0.0208                                   & 0.0523                                   & 0.0364                                   & 0.0693                                   & 0.0419                                   & 0.0915                                   & 0.0474                                   \\
\multicolumn{1}{c|}{}                                   & C\textsc{a}F\textsc{e}                                      & 0.0235                                   & 0.0150                                   & 0.0384                                   & 0.0198                                   & 0.0614                                   & 0.0257                                   & 0.0598                                   & 0.0442                                   & 0.0743                                   & 0.0489                                   & 0.0967                                   & 0.0545                                   \\
\multicolumn{1}{c|}{}                                   & LLM4SBR                                   & 0.0164                                   & 0.0091                                   & 0.0248                                   & 0.0119                                   & 0.0405                                   & 0.0158                                   & 0.0578                                   & 0.0419                                   & 0.0752                                   & 0.0476                                   & 0.1002                                   & 0.0539                                   \\
\multicolumn{1}{c|}{}                                   & LLM-ESR                                   & 0.0191                                   & 0.0126                                   & 0.0264                                   & 0.0149                                   & 0.0380                                   & 0.0178                                   & 0.0487                                   & 0.0343                                   & 0.0698                                   & 0.0410                                   & 0.0973                                   & 0.0480                                   \\
\multicolumn{1}{c|}{}                                   & RLMRec-Con                                & 0.0224                             & 0.0128                                   & 0.0376                             & 0.0176                             & 0.0585                             & 0.0228                             & 0.0550                                   & 0.0386                                   & 0.0752                                   & 0.0451                                   & 0.0978                                   & 0.0507                                   \\
\multicolumn{1}{c|}{}                                   & RLMRec-Gen                                & 0.0210                                   & 0.0135                             & 0.0329                                   & 0.0173                                   & 0.0525                                   & 0.0223                                   & 0.0636                             & 0.0470                             & 0.0832                             & 0.0534                             & 0.1056                             & 0.0590                             \\
\multicolumn{1}{c|}{}                                   & ProfileRec                                & {\ul 0.0293}                                   & {\ul 0.0183}                             & {\ul 0.0503}                                   & {\ul 0.0250}                                   & {\ul 0.0763}                                   & {\ul 0.0315}                                   & {\ul 0.0672}                             & {\ul 0.0490}                             & {\ul 0.0901}                             & {\ul 0.0565}                             & {\ul 0.1163}                             & {\ul 0.0630}                             \\
\multicolumn{1}{c|}{}& \cellcolor[HTML]{D9D9D9}SPRINT & \cellcolor[HTML]{D9D9D9}\textbf{0.0341*} & \cellcolor[HTML]{D9D9D9}\textbf{0.0224*} & \cellcolor[HTML]{D9D9D9}\textbf{0.0548*} & \cellcolor[HTML]{D9D9D9}\textbf{0.0290*} & \cellcolor[HTML]{D9D9D9}\textbf{0.0796*} & \cellcolor[HTML]{D9D9D9}\textbf{0.0352*} & \cellcolor[HTML]{D9D9D9}0.0722 & \cellcolor[HTML]{D9D9D9}0.0521 & \cellcolor[HTML]{D9D9D9}0.0936 & \cellcolor[HTML]{D9D9D9}0.0590 & \cellcolor[HTML]{D9D9D9}0.1196 & \cellcolor[HTML]{D9D9D9}0.0655                                              \\
\multicolumn{1}{c|}{\multirow{-16}{*}{\textbf{Book}}}   & \cellcolor[HTML]{D9D9D9}SPRINT+LE  & \cellcolor[HTML]{D9D9D9}0.0164           & \cellcolor[HTML]{D9D9D9}0.0110           & \cellcolor[HTML]{D9D9D9}0.0265           & \cellcolor[HTML]{D9D9D9}0.0142           & \cellcolor[HTML]{D9D9D9}0.0378           & \cellcolor[HTML]{D9D9D9}0.0170           & \cellcolor[HTML]{D9D9D9}\textbf{0.0749*}           & \cellcolor[HTML]{D9D9D9}\textbf{0.0533*}           & \cellcolor[HTML]{D9D9D9}\textbf{0.0971*}           & \cellcolor[HTML]{D9D9D9}\textbf{0.0605*}           & \cellcolor[HTML]{D9D9D9}\textbf{0.1250*}           & \cellcolor[HTML]{D9D9D9}\textbf{0.0676*}
\\ \bottomrule
\vspace{-0.1cm}
\end{tabular}
}\label{tab:maintable}
\end{table*}

%% file: sections/051Experiment_result.tex
\input{tables/tab_llama_fix2}
\subsection{Performance Comparison}\label{sub:performance_comparision}
Table \ref{tab:maintable} presents the performance comparison.
Overall, the proposed framework consistently outperforms all baselines across all datasets and metrics.
This improvement stems from three key factors.

First, in contrast to implicit intent modeling (i.e., ELCRec, MiaSRec), \proposed explicitly infers intents via LLM reasoning, providing semantically meaningful guidance to the SBR model.
%
Second, compared to explicit intent modeling (i.e., FAPAT, C\textsc{a}F\textsc{e}) that directly utilize raw item features as pseudo-intents,
\proposed generates semantically rich and nuanced session-level intents by leveraging the reasoning capabilities and world knowledge of LLMs.
%
Third, unlike LLM-enhanced CRMs (i.e., LLM4SBR, LLM-ESR, RLMRec, ProfileRec), \proposed refines and validates intents based on recommendation performance via a P\&C loop coupled with a global intent pool, ensuring highly reliable and informative intents.
%


Interestingly, \proposed\ with LE outperforms \proposed\ only when paired with BERT4Rec.
We speculate that LLM embeddings help alleviate data sparsity in scenarios where BERT’s bidirectional modeling struggles, such as sparse interactions in Beauty, very short sessions in Yelp (average length $<4$), and the large item space in Book.
In contrast, incorporating LE into SASRec degrades performance on Book. We attribute this to Book’s relatively longer sequences combined with SASRec’s unidirectional modeling, which makes the model more susceptible to noise introduced by LLM embeddings.
This degradation trend is consistently observed in other LE-equipped baselines, including LLM4SBR and LLM-ESR.
This highlights the need for further investigation into how the effectiveness of LLM embeddings depends on model architecture and data characteristics, which we leave for future work.

\smallsection{Results with open-source LLM.}
Table~\ref{tab:llama} presents the results of LLM-enhanced methods on Beauty using an open-source LLM and its embeddings.
Overall, \proposed significantly outperforms all baselines,
showing similar patterns to Table~\ref{tab:maintable}.
This result highlights the generalizability of our method across various LLMs.
Interestingly, models equipped with LE (i.e., LLM-ESR, \proposed+LE) perform poorly.
We attribute this to the representational mismatch between the high-dimensional LLaMA embeddings (8192) and the much lower-dimensional SBR model (64), indicating that more sophisticated alignment techniques may be required.


\subsection{Efficiency Analysis}\label{sub:efficiency_analysis}
As real-world systems process massive numbers of sessions, generating intents via LLMs for every session is highly inefficient for both training and inference.
In this section, we analyze the efficiency of \proposed. 
We use SASRec as the backbone SBR model. 

\begin{figure}[t!]
  \includegraphics[width=1.0\columnwidth]{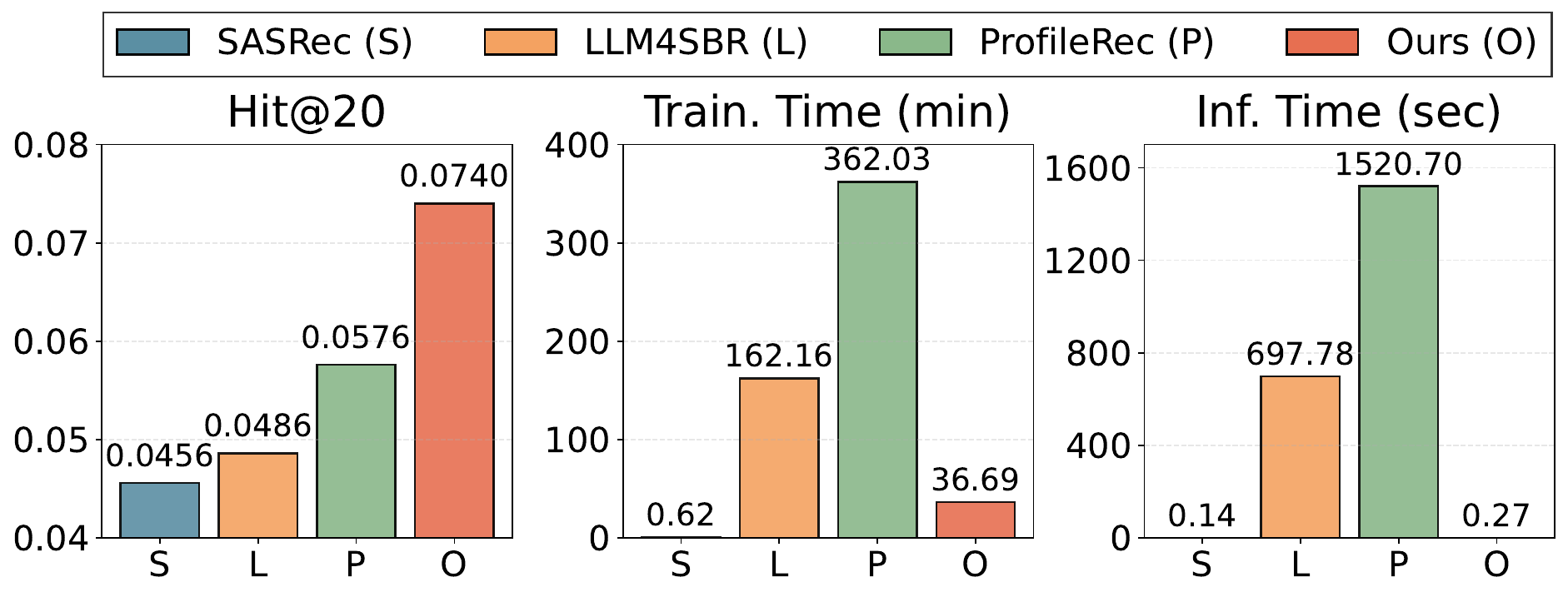}
    \caption{Comparison of performance and efficiency on Beauty. "Train. Time" denotes the total time, including LLM-based user profiling for session contexts (e.g., profiles or intents) and subsequent SBR model training. "Inf. Time" denotes the inference latency for the entire test set.}
  \label{fig:perfomrance_efficiency}
  \vspace{-0.1cm}
\end{figure}

\smallsection{Accuracy–efficiency trade-off.}
Figure~\ref{fig:perfomrance_efficiency} shows the trade-off between performance and training/inference time.
\proposed achieves superior accuracy with significantly reduced training time compared to LLM-based user profiling methods, while maintaining inference latency comparable to the LLM-free SASRec. 
We attribute this favorable balance to two key design choices:
(1) informative and reliable intents enhance the SBR model's accuracy, and 
(2) selective LLM-based intent generation for uncertain sessions, combined with a lightweight intent predictor, which significantly reduces computational overhead.
As a result, \proposed effectively addresses both uninformative outputs (\textbf{C1}) and scalability limitations (\textbf{C2}).


\smallsection{Comparison with an LLM-as-SBR model.}
A key advantage of \proposed is its elimination of LLM dependency at inference time.
To highlight this efficiency, we compare \proposed with PO4ISR~\cite{PO4ISR}, a state-of-the-art LLM-as-SBR method that performs both intent generation and recommendation using LLMs during inference.
Following~\cite{PO4ISR}, we use 20 candidate items, consisting of 19 negative items from a pretrained SASRec and one ground-truth item.
Table~\ref{tab:inference_efficiency} shows that \proposed\! with LE significantly outperforms PO4ISR while drastically reducing inference time.
Notably, even without LLM embeddings, \proposed shows highly competitive performance on all datasets except Beauty.
These results highlight the efficiency of \proposed, which uses only a lightweight intent predictor at test time and requires no LLM inference, enabling scalable deployment.

\input{tables/tab_inference_efficiency_fix}

\subsection{Design Choice Analysis}\label{sub:study_of_SPRINT}

\subsubsection{\textbf{Ablation study}}\label{subsub:ablation_study}
Table~\ref{tab:abl} presents ablation results on Beauty.
Notably, using all proposed components yields the best performance, confirming their efficacy.
Specifically, incorporating (A) LLM-generated intents shows a significant improvement compared to when excluding this component.
The inclusion of (B) the P\&C loop with the GIP further enhances performance, showing its capability to generate reliable and informative intents. 
Note that the GIP cannot be meaningfully ablated; without it, the unbounded intent space renders intent prediction and fusion in Stage~2 infeasible.
Finally, both enrichment variants, (C) self-prediction and (D) neighbor-aggregated prediction, show marked improvements, indicating that our strategy effectively performs intent label enrichment by leveraging both self-confidence and collaborative signals.

\begin{figure}[t!]
  \centering
  \resizebox{0.9\columnwidth}{!}{%
    \includegraphics{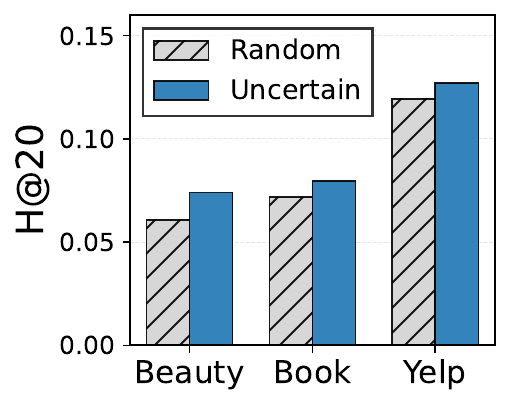}%
    \hspace{1pt}%
    \includegraphics{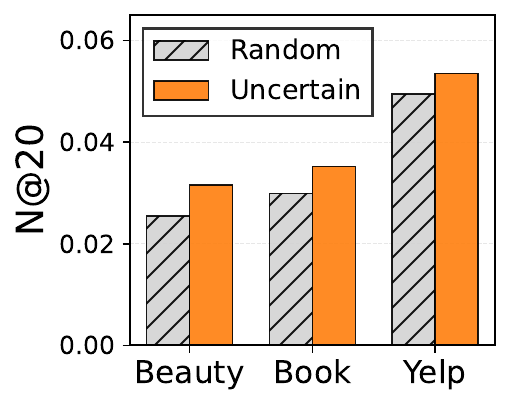}%
  }
  \caption{Random vs. uncertainty-aware session selection.}
  \label{fig:random_vs_uncertain}
  \vspace{-0.1cm}
\end{figure}

Figure~\ref{fig:random_vs_uncertain} compares the performance of our uncertainty-aware session selection against a random selection.
Our approach consistently outperforms the random selection across all datasets.
This result confirms that prioritizing high-uncertainty sessions is more effective than random sampling, as it strategically applies LLM calls to the cases where nuanced reasoning is most needed.

\input{tables/tab_abl_final}

\vspace{-0cm}
\subsubsection{\textbf{Hyperparameter study}}\label{subsub:hyperparameter_study}
\begin{figure}[t]
  \centering
  \resizebox{1.0\columnwidth}{!}{%
    \includegraphics{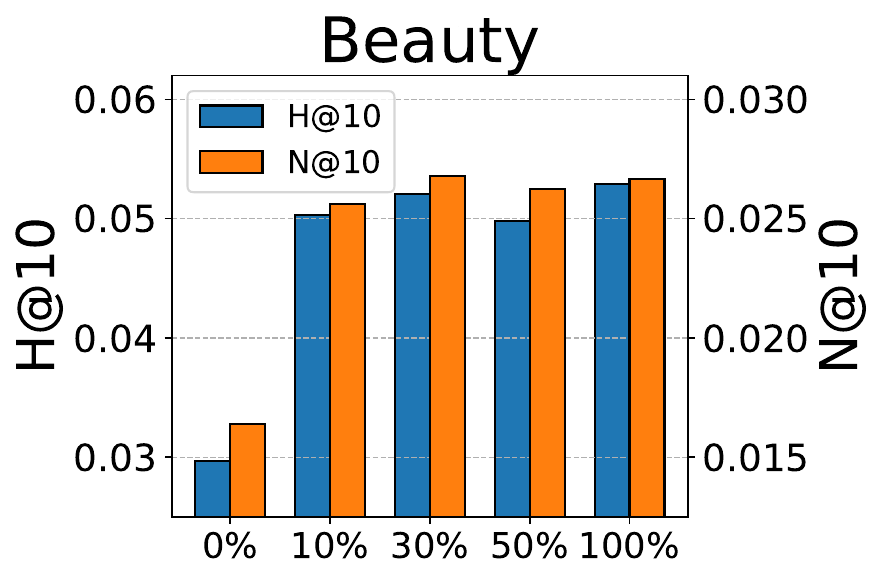}%
    \hspace{1pt}%
    \includegraphics{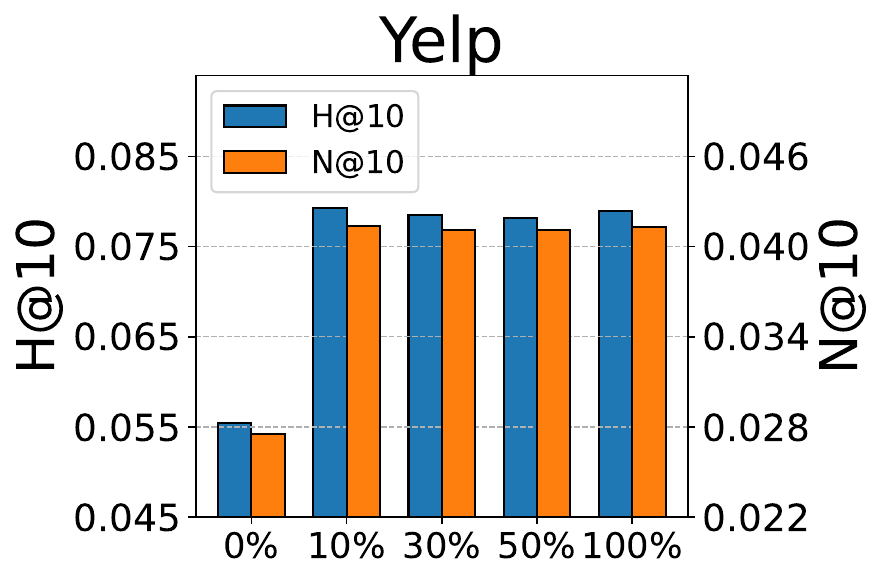}%
  }
  \caption{Results with varying amounts of LLM intents.}
  \label{fig:intent_percentage}
    \vspace{-0.3cm}
\end{figure}

We provide analysis to guide the hyperparameter selection of \proposed.
First, we analyze the effect of the intent supervision ratio, i.e., the proportion of sessions for which LLM-generated intents are used.
In our main experiments, we generate LLM-based intents only for the top 10\% most uncertain sessions (\S~\ref{subsub:uncertainty_select}).
As shown in Figure~\ref{fig:intent_percentage}, \proposed achieves competitive performance even with partial intent supervision and significantly outperforms the 0\% setting.
We attribute this robustness to the synergy between validated intents from the P\&C loop and collaborative intent enrichment, which enables effective propagation of high-quality intent information.
Therefore, \proposed effectively generalizes learned intent patterns to the remaining sessions without explicit LLM intent label supervision.

\begin{figure}[t!]
  \centering
  \begin{subfigure}[c]{0.4\columnwidth}
    \centering
    \includegraphics[width=\linewidth]{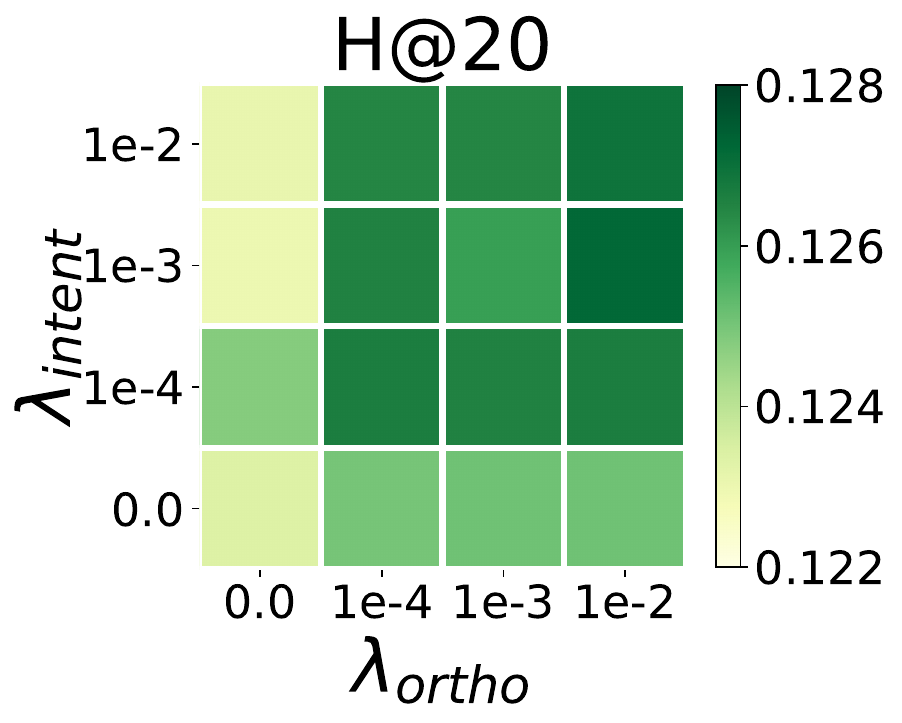}
  \end{subfigure}%
  \begin{subfigure}[c]{0.59\columnwidth}
    \centering
    \includegraphics[width=\linewidth]{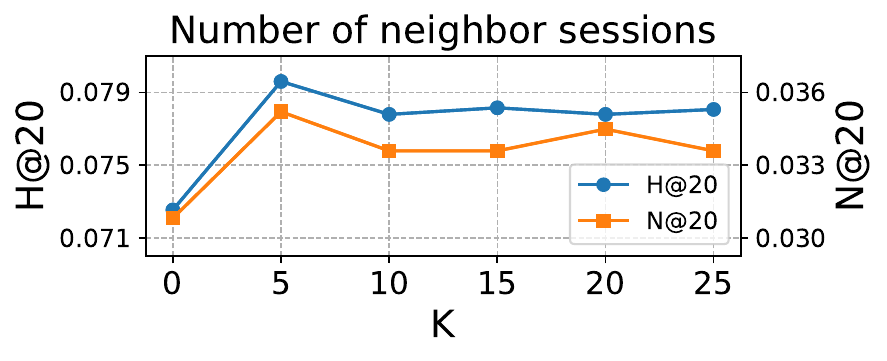}
  \end{subfigure}
  \caption{Impact of $\lambda_{\text{intent}}, \lambda_{\text{ortho}}$, and top-$K$ neighbors.} 
  \label{fig:lambda_numk}
\end{figure}

Figure~\ref{fig:lambda_numk} (left) provides the results of varying $\lambda_{\text{intent}} \text{ and } \lambda_{\text{ortho}}$ on Yelp.
Setting either weight to 0.0 leads to degraded performance, confirming the necessity of both loss terms.
Meanwhile, non-zero values consistently improve performance, indicating stable tuning with low sensitivity to exact values.
Finally, Figure~\ref{fig:lambda_numk} (right) examines the impact of the number of top-$K$ neighbor sessions (\cref{subsub:int_enrich}) on Book, where $K=0$ indicates no neighbor usage.
Performance drops sharply at $K=0$, highlighting the importance of neighbor-based intent enrichment.
Performance peaks at $K=5$, suggesting that a small set of relevant neighbors is sufficient for effective enrichment.






\subsection{Case Study}\label{sub:case_study}
\begin{figure*}[t]
  \centering
  \resizebox{0.95\linewidth}{!}{%
    \includegraphics{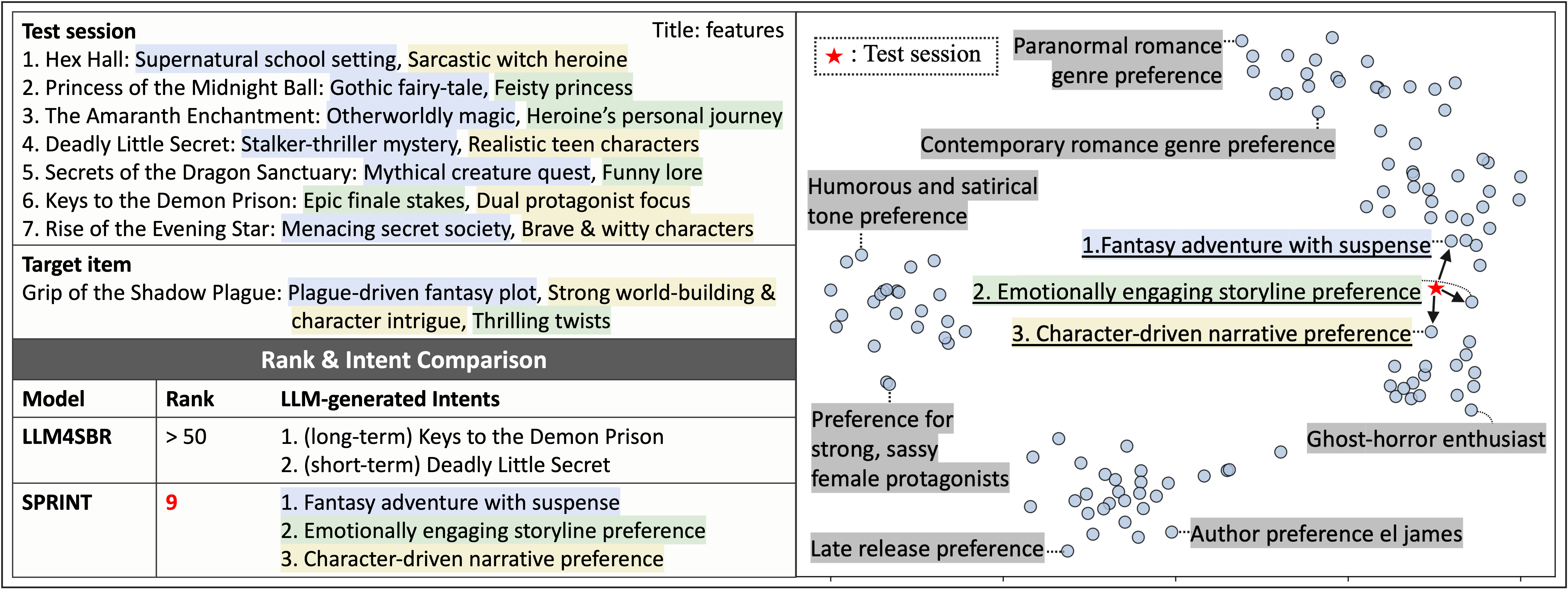}
  }
  \caption{Case study on the Book dataset. (Left) Rank and intent comparison for a test session. 
  Same-colored boxes indicate similar semantic meanings across the test session, target item, and inferred intents.
  (Right) t-SNE visualization of the learned intent embeddings. The red star denotes the embedding of the test session shown on the left. 
  }
  \label{fig:case_study_book}
  \vspace{-0.3cm}
\end{figure*}

Figure~\ref{fig:case_study_book} presents a case study.
We compare \proposed with LLM4SBR, a state-of-the-art LLM-enhanced SBR method that explicitly generates interpretable intents.
Note that other LLM-based baselines focus on general user profiling in the form of embeddings or personas rather than explicit intent modeling, making direct intent-level comparison via visualization infeasible.
Our inferred intents (e.g., ``Fantasy adventure with suspense'', ``Emotionally engaging storyline preference'') capture the core characteristics of both the test session and the target item.
In contrast, LLM4SBR derives intents directly from item titles (e.g., ``Keys to the Demon Prison'', ``Deadly Little Secret''), resulting in vaguer and less informative representations.
Consequently, our more informative intents lead to improved ranking performance.
The t-SNE visualization further shows that the test session embedding is closely aligned with its inferred intents, highlighting the effectiveness of our multi-intent inference process.
Overall, this supports that \proposed improves both accuracy and explainability through high-quality~intents.
\vspace{-0cm}

%% file: tables/tab_llama_fix2.tex
\begin{table}[t]
\caption{Performance comparison of LLM‑enhanced CRMs on the Beauty dataset using Llama‑3.3‑70B‑Instruct.}
\centering
\renewcommand{\arraystretch}{0.8}
\resizebox{\columnwidth}{!}{%
\begin{tabular}{c|l|cccccc}
\toprule
\multicolumn{2}{c|}{\textbf{Method}} & \textbf{H@5} & \textbf{N@5} & \textbf{H@10} & \textbf{N@10} & \textbf{H@20} & \textbf{N@20} \\ \midrule\midrule
\multirow{7}{*}{\rotatebox[origin=c]{90}{\textbf{SASRec}}} 
 & LLM4SBR & 0.0159 & 0.0107 & 0.0293 & 0.0149 & 0.0435 & 0.0185 \\
 & LLM-ESR & 0.0073 & 0.0038 & 0.0099 & 0.0046 & 0.0159 & 0.0061 \\
 & RLMRec-Con & 0.0219 & 0.0133 & 0.0357 & 0.0176 & 0.0602 & 0.0239 \\
 & RLMRec-Gen & 0.0207 & 0.0125 & 0.0301 & 0.0155 & 0.0546 & 0.0217 \\
 & ProfileRec & {\ul 0.0258} & {\ul 0.0166} & {\ul 0.0375} & {\ul 0.0204} & {\ul 0.0620} & {\ul 0.0264} \\
 & \cellcolor[HTML]{D9D9D9}SPRINT & \cellcolor[HTML]{D9D9D9}\textbf{0.0314*} & \cellcolor[HTML]{D9D9D9}\textbf{0.0194*} & \cellcolor[HTML]{D9D9D9}\textbf{0.0512*} & \cellcolor[HTML]{D9D9D9}\textbf{0.0257*} & \cellcolor[HTML]{D9D9D9}\textbf{0.0744*} & \cellcolor[HTML]{D9D9D9}\textbf{0.0316*} \\
 & \cellcolor[HTML]{D9D9D9}SPRINT+LE & \cellcolor[HTML]{D9D9D9}0.0076 & \cellcolor[HTML]{D9D9D9}0.0038 & \cellcolor[HTML]{D9D9D9}0.0140 & \cellcolor[HTML]{D9D9D9}0.0058 & \cellcolor[HTML]{D9D9D9}0.0226 & \cellcolor[HTML]{D9D9D9}0.0080 \\ \midrule
\multirow{7}{*}{\rotatebox[origin=c]{90}{\textbf{BERT4Rec}}} 
 & LLM4SBR & 0.0116 & 0.0070 & 0.0220 & 0.0103 & 0.0447 & 0.0160 \\
 & LLM-ESR & 0.0056 & 0.0028 & 0.0112 & 0.0046 & 0.0194 & 0.0066 \\
 & RLMRec-Con & 0.0232 & 0.0153 & 0.0340 & 0.0187 & 0.0572 & 0.0245 \\
 & RLMRec-Gen & {\ul 0.0250} & {\ul 0.0169} & {\ul 0.0374} & {\ul 0.0209} & 0.0568 & 0.0258 \\
 & ProfileRec & 0.0245 & 0.0167 & 0.0361 & 0.0205 & {\ul 0.0688} & {\ul 0.0285} \\
 & \cellcolor[HTML]{D9D9D9}SPRINT & \cellcolor[HTML]{D9D9D9}\textbf{0.0258} & \cellcolor[HTML]{D9D9D9}\textbf{0.0170} & \cellcolor[HTML]{D9D9D9}\textbf{0.0417*} & \cellcolor[HTML]{D9D9D9}\textbf{0.0221} & \cellcolor[HTML]{D9D9D9}\textbf{0.0735*} & \cellcolor[HTML]{D9D9D9}\textbf{0.0301*} \\
  
 & \cellcolor[HTML]{D9D9D9}SPRINT+LE &  \cellcolor[HTML]{D9D9D9}0.0086 & \cellcolor[HTML]{D9D9D9}0.0065 & \cellcolor[HTML]{D9D9D9}0.0118 & \cellcolor[HTML]{D9D9D9}0.0076 & \cellcolor[HTML]{D9D9D9}0.0183 & \cellcolor[HTML]{D9D9D9}0.0091 \\ \bottomrule
\end{tabular}%
}
\label{tab:llama}
\end{table}

%% file: tables/tab_inference_efficiency_fix.tex
\begin{table}[t] \vspace{-0.1cm}
\caption{Recommendation accuracy and total inference time of PO4ISR (LLM‑as‑SBR) vs. \proposed (LLM-enhanced SBR).
Total inference time refers to the cumulative time required to generate predictions for all test sessions.}
\centering
\renewcommand{\arraystretch}{0.8}
\resizebox{\columnwidth}{!}{%
\begin{tabular}{c|l|ccccc}
\toprule
\textbf{Dataset}        & \textbf{Method}   & \textbf{H@5}    & \textbf{N@5}    & \textbf{H@10}   & \textbf{N@10}   & \textbf{Inf. Time} \\ \midrule\midrule
\multirow{3}{*}{\textbf{Beauty}} & PO4ISR            & 0.3996& \textbf{0.2872}& 0.5491& 0.3351& 53.5 min      \\
                        & SPRINT & 0.3372& 0.2513& 0.4310& 0.2814& 0.12 s        \\
                        & SPRINT+LE  & \textbf{0.4293}& 0.2769& \textbf{0.6718}& \textbf{0.3550}& 0.31 s        \\
\midrule
\multirow{3}{*}{\textbf{Yelp}}   & PO4ISR            & 0.2764& 0.1171& 0.4877& 0.2447& 371.5 min     \\
                        & SPRINT & 0.6015& 0.4838& 0.7158& 0.5206& 0.23 s        \\
                        & SPRINT+LE  & \textbf{0.6368}& \textbf{0.5033}& \textbf{0.7600}& \textbf{0.5430}& 0.57 s        \\
\midrule
\multirow{3}{*}{\textbf{Book}}   & PO4ISR            & 0.3892& 0.3025& 0.5124& 0.3419& 161.9 min     \\
                        & SPRINT & 0.4280& 0.3320& 0.5608& 0.3749& 0.14 s       \\
                        & SPRINT+LE  & \textbf{0.5246}& \textbf{0.3811}& \textbf{0.6972}& \textbf{0.4366}& 0.39 s    
\\ \bottomrule
\end{tabular}
}
\label{tab:inference_efficiency}
\vspace{-0.3cm}
\end{table}

%% file: tables/tab_abl_final.tex
\begin{table}[t] \vspace{-0.1cm}
\centering
\caption{Ablation study: (A) LLM-generated intents, (B) P\&C loop with GIP, (C) self-enrichment, and (D) neighbor-enrichment.
(C) and (D) correspond to intent label enrichment using self-prediction and neighbor-prediction only.}
\renewcommand{\arraystretch}{0.4} 
\resizebox{0.9\columnwidth}{!}{%
\begin{tabular}{cccc|cccc}
\toprule
\textbf{(A)} & \textbf{(B)} & \textbf{(C)} & \textbf{(D)} & \textbf{H@5} & \textbf{N@5} & \textbf{H@20} & \textbf{N@20} \\ 
\midrule\midrule
$\times$ & $\times$ & $\times$ & $\times$ & 0.0198 & 0.0132 & 0.0456 & 0.0203 \\
$\checkmark$ & & & & 0.0271 & 0.0164 & 0.0585 & 0.0257 \\
$\checkmark$ & $\checkmark$ & & & 0.0310 & 0.0192 & 0.0619 & 0.0280 \\
$\checkmark$ & $\checkmark$ & $\checkmark$ & & 0.0323 & 0.0195 & 0.0632 & 0.0282 \\
$\checkmark$ & $\checkmark$ & & $\checkmark$ & 0.0327 & 0.0197 & 0.0675 & 0.0295 \\
$\checkmark$ & $\checkmark$ & $\checkmark$ & $\checkmark$ & \textbf{0.0331} & \textbf{0.0201} & \textbf{0.0740} & \textbf{0.0316} \\
\bottomrule
\end{tabular}
}
\vspace{-0.185cm}
\label{tab:abl}
\end{table}




%% file: sections/070Conclusion.tex
In this paper, we propose \proposed, a two-stage framework designed to address two unique challenges in LLM-based intent profiling for SBR.
For \textbf{C1}, \proposed iteratively refines and validates LLM-generated intents via the P\&C loop with the global intent pool, while further enriching them using collaborative signals.
For \textbf{C2}, \proposed strategically invokes LLMs via uncertainty-aware session selection and employs a lightweight, LLM-free intent predictor.
Our experiments show that \proposed not only outperforms the baselines but also provides meaningful interpretability.
We believe \proposed provides a practical path toward the scalable and effective deployment of LLM-enhanced SBR systems in real-world scenarios.
Future directions include adapting \proposed to continual settings where user interests evolve over non-stationary data streams~\cite{lee2024continual, lee2026capturing}.

%% file: acmart.bib
@inproceedings{kang2024improving,
  title={Improving retrieval in theme-specific applications using a corpus topical taxonomy},
  author={Kang, SeongKu and Agarwal, Shivam and Jin, Bowen and Lee, Dongha and Yu, Hwanjo and Han, Jiawei},
  booktitle={Proceedings of the ACM Web Conference 2024},
  pages={1497--1508},
  year={2024}
}

@inproceedings{lee2026capturing,
  title={Capturing User Interests from Data Streams for Continual Sequential Recommendation},
  author={Lee, Gyuseok and Yoo, Hyunsik and Hwang, Junyoung and Kang, SeongKu and Yu, Hwanjo},
  booktitle={Proceedings of the Nineteenth ACM International Conference on Web Search and Data Mining},
  pages={313--323},
  year={2026}
}

@inproceedings{lee2025collaborative,
  title={Collaborative Diffusion Model for Recommender System},
  author={Lee, Gyuseok and Zhu, Yaochen and Yu, Hwanjo and Zhou, Yao and Li, Jundong},
  booktitle={Companion Proceedings of the ACM on Web Conference 2025},
  pages={1091--1095},
  year={2025}
}

@inproceedings{lee2024continual,
  title={Continual collaborative distillation for recommender system},
  author={Lee, Gyuseok and Kang, SeongKu and Kweon, Wonbin and Yu, Hwanjo},
  booktitle={Proceedings of the 30th ACM SIGKDD Conference on Knowledge Discovery and Data Mining},
  pages={1495--1505},
  year={2024}
}

@inproceedings{kweon2025topic,
  title={Topic Coverage-based Demonstration Retrieval for In-Context Learning},
  author={Kweon, Wonbin and Kang, SeongKu and Tian, Runchu and Jiang, Pengcheng and Han, Jiawei and Yu, Hwanjo},
  booktitle={Proceedings of the 2025 Conference on Empirical Methods in Natural Language Processing},
  pages={19911--19923},
  year={2025}
}

@inproceedings{kweon2024doubly,
  title={Doubly calibrated estimator for recommendation on data missing not at random},
  author={Kweon, Wonbin and Yu, Hwanjo},
  booktitle={Proceedings of the ACM Web Conference 2024},
  pages={3810--3820},
  year={2024}
}

@article{kweon2025pairsem,
  title={PairSem: LLM-Guided Pairwise Semantic Matching for Scientific Document Retrieval},
  author={Kweon, Wonbin and Tian, Runchu and Kang, SeongKu and Jiang, Pengcheng and Lu, Zhiyong and Han, Jiawei and Yu, Hwanjo},
  journal={arXiv preprint arXiv:2510.09897},
  year={2025}
}

@inproceedings{yoo2025embracing,
  title={Embracing plasticity: Balancing stability and plasticity in continual recommender systems},
  author={Yoo, Hyunsik and Kang, SeongKu and Qiu, Ruizhong and Xu, Charlie and Wang, Fei and Tong, Hanghang},
  booktitle={Proceedings of the 48th International ACM SIGIR conference on research and development in Information Retrieval},
  pages={2092--2101},
  year={2025}
}

@inproceedings{yoo2025continual,
  title={Continual Low-Rank Adapters for LLM-based Generative Recommender Systems},
  author={Yoo, Hyunsik and Li, Ting-Wei and Kang, SeongKu and Liu, Zhining and Xu, Charlie and Qi, Qilin and Tong, Hanghang},
  booktitle={International Conference on Learning Representations (ICLR)},
  year={2026}
}

@inproceedings{yoo2024ensuring,
  title={Ensuring user-side fairness in dynamic recommender systems},
  author={Yoo, Hyunsik and Zeng, Zhichen and Kang, Jian and Qiu, Ruizhong and Zhou, David and Liu, Zhining and Wang, Fei and Xu, Charlie and Chan, Eunice and Tong, Hanghang},
  booktitle={Proceedings of the ACM web conference 2024},
  pages={3667--3678},
  year={2024}
}

@article{Hit,
  title={Item-based top-n recommendation algorithms},
  author={Deshpande, Mukund and Karypis, George},
  journal={ACM Transactions on Information Systems (TOIS)},
  volume={22},
  number={1},
  pages={143--177},
  year={2004},
  publisher={ACM New York, NY, USA}
}

@article{NDCG,
  title={Cumulated gain-based evaluation of IR techniques},
  author={J{\"a}rvelin, Kalervo and Kek{\"a}l{\"a}inen, Jaana},
  journal={ACM Transactions on Information Systems (TOIS)},
  volume={20},
  number={4},
  pages={422--446},
  year={2002},
  publisher={ACM New York, NY, USA}
}

@article{sigmoid,
  title={Learning representations by back-propagating errors},
  author={Rumelhart, David E and Hinton, Geoffrey E and Williams, Ronald J},
  journal={nature},
  volume={323},
  number={6088},
  pages={533--536},
  year={1986},
  publisher={Nature Publishing Group UK London}
}

@incollection{softmax,
  title={Probabilistic interpretation of feedforward classification network outputs, with relationships to statistical pattern recognition},
  author={Bridle, John S},
  booktitle={Neurocomputing: Algorithms, architectures and applications},
  pages={227--236},
  year={1990},
  publisher={Springer}
}

@inproceedings{mcauley2015image,
  title={Image-based recommendations on styles and substitutes},
  author={McAuley, Julian and Targett, Christopher and Shi, Qinfeng and Van Den Hengel, Anton},
  booktitle={Proceedings of the 38th international ACM SIGIR conference on research and development in information retrieval},
  pages={43--52},
  year={2015}
}

@misc{yelp_open_dataset,
  author       = {{Yelp}},
  title        = {Yelp Open Dataset},
  howpublished = {\url{https://business.yelp.com/data/resources/open-dataset/}},
  note         = {Accessed: 2026-01-20}
}

@inproceedings{wang2025intent,
  title={Intent representation learning with large language model for recommendation},
  author={Wang, Yu and Sang, Lei and Zhang, Yi and Zhang, Yiwen},
  booktitle={Proceedings of the 48th International ACM SIGIR Conference on Research and Development in Information Retrieval},
  pages={1870--1879},
  year={2025}
}

@inproceedings{bougie2025simuser,
  title={Simuser: Simulating user behavior with large language models for recommender system evaluation},
  author={Bougie, Nicolas and Watanabe, Narimawa},
  booktitle={Proceedings of the 63rd Annual Meeting of the Association for Computational Linguistics (Volume 6: Industry Track)},
  pages={43--60},
  year={2025}
}

@article{wang2022learning,
  title={Learning persona-driven personalized sentimental representation for review-based recommendation},
  author={Wang, Peipei and Li, Lin and Wang, Ru and Zheng, Xinhao and He, Jiaxi and Xu, Guandong},
  journal={Expert Systems with Applications},
  volume={203},
  pages={117317},
  year={2022},
  publisher={Elsevier}
}

@article{li2025ctrl,
  title={Ctrl: Connect collaborative and language model for ctr prediction},
  author={Li, Xiangyang and Chen, Bo and Hou, Lu and Tang, Ruiming},
  journal={ACM Transactions on Recommender Systems},
  volume={4},
  number={2},
  pages={1--23},
  year={2025},
  publisher={ACM New York, NY}
}

@inproceedings{lyu2024llm,
  title={Llm-rec: Personalized recommendation via prompting large language models},
  author={Lyu, Hanjia and Jiang, Song and Zeng, Hanqing and Xia, Yinglong and Wang, Qifan and Zhang, Si and Chen, Ren and Leung, Chris and Tang, Jiajie and Luo, Jiebo},
  booktitle={Findings of the Association for Computational Linguistics: NAACL 2024},
  pages={583--612},
  year={2024}
}

@inproceedings{wang2024rdrec,
  title={RDRec: Rationale Distillation for LLM-based Recommendation},
  author={Wang, Xinfeng and Cui, Jin and Suzuki, Yoshimi and Fukumoto, Fumiyo},
  booktitle={Proceedings of the 62nd Annual Meeting of the Association for Computational Linguistics (Volume 2: Short Papers)},
  pages={65--74},
  year={2024}
}

@inproceedings{wang2025large,
  title={Large Language Model driven Policy Exploration for Recommender Systems},
  author={Wang, Jie and Karatzoglou, Alexandros and Arapakis, Ioannis and Jose, Joemon M},
  booktitle={Proceedings of the Eighteenth ACM International Conference on Web Search and Data Mining},
  pages={107--116},
  year={2025}
}

@inproceedings{liu2025llmemb,
  title={Llmemb: Large language model can be a good embedding generator for sequential recommendation},
  author={Liu, Qidong and Wu, Xian and Wang, Wanyu and Wang, Yejing and Zhu, Yuanshao and Zhao, Xiangyu and Tian, Feng and Zheng, Yefeng},
  booktitle={Proceedings of the AAAI Conference on Artificial Intelligence},
  volume={39},
  number={11},
  pages={12183--12191},
  year={2025}
}

@inproceedings{hu2024enhancing,
  title={Enhancing sequential recommendation via llm-based semantic embedding learning},
  author={Hu, Jun and Xia, Wenwen and Zhang, Xiaolu and Fu, Chilin and Wu, Weichang and Huan, Zhaoxin and Li, Ang and Tang, Zuoli and Zhou, Jun},
  booktitle={Companion Proceedings of the ACM Web Conference 2024},
  pages={103--111},
  year={2024}
}

@inproceedings{shi2025you,
  title={You are what you bought: Generating customer personas for e-commerce applications},
  author={Shi, Yimin and Fei, Yang and Zhang, Shiqi and Wang, Haixun and Xiao, Xiaokui},
  booktitle={Proceedings of the 48th International ACM SIGIR Conference on Research and Development in Information Retrieval},
  pages={1810--1819},
  year={2025}
}

@inproceedings{song2019session,
  title={Session-based social recommendation via dynamic graph attention networks},
  author={Song, Weiping and Xiao, Zhiping and Wang, Yifan and Charlin, Laurent and Zhang, Ming and Tang, Jian},
  booktitle={Proceedings of the Twelfth ACM international conference on web search and data mining},
  pages={555--563},
  year={2019}
}

@article{bruch2023analysis,
  title={An analysis of fusion functions for hybrid retrieval},
  author={Bruch, Sebastian and Gai, Siyu and Ingber, Amir},
  journal={ACM Transactions on Information Systems},
  volume={42},
  number={1},
  pages={1--35},
  year={2023},
  publisher={ACM New York, NY}
}

@inproceedings{cormack2009reciprocal,
  title={Reciprocal rank fusion outperforms condorcet and individual rank learning methods},
  author={Cormack, Gordon V and Clarke, Charles LA and Buettcher, Stefan},
  booktitle={Proceedings of the 32nd international ACM SIGIR conference on Research and development in information retrieval},
  pages={758--759},
  year={2009}
}

@inproceedings{ji2023towards,
  title={Towards mitigating LLM hallucination via self reflection},
  author={Ji, Ziwei and Yu, Tiezheng and Xu, Yan and Lee, Nayeon and Ishii, Etsuko and Fung, Pascale},
  booktitle={Findings of the Association for Computational Linguistics: EMNLP 2023},
  pages={1827--1843},
  year={2023}
}

@inproceedings{ProfileRec,
  title={Enhancing Recommendation with Reliable Multi-profile Alignment and Collaborative-aware Contrastive Learning},
  author={Liu, Yibin and Zhang, Jianyu and Li, Shijian},
  booktitle={Proceedings of the 34th ACM International Conference on Information and Knowledge Management},
  pages={1936--1946},
  year={2025}
}

@inproceedings{jia2025learn,
  title={LEARN: Knowledge Adaptation from Large Language Model to Recommendation for Practical Industrial Application},
  author={Jia, Jian and Wang, Yipei and Li, Yan and Chen, Honggang and Bai, Xuehan and Liu, Zhaocheng and Liang, Jian and Chen, Quan and Li, Han and Jiang, Peng and others},
  booktitle={Proceedings of the AAAI Conference on Artificial Intelligence},
  volume={39},
  number={11},
  pages={11861--11869},
  year={2025}
}

@inproceedings{wei2024llmrec,
  title={Llmrec: Large language models with graph augmentation for recommendation},
  author={Wei, Wei and Ren, Xubin and Tang, Jiabin and Wang, Qinyong and Su, Lixin and Cheng, Suqi and Wang, Junfeng and Yin, Dawei and Huang, Chao},
  booktitle={Proceedings of the 17th ACM international conference on web search and data mining},
  pages={806--815},
  year={2024}
}

@inproceedings{liu2025large,
  title={Large Language Model Enhanced Recommender Systems: Methods, Applications and Trends},
  author={Liu, Qidong and Zhao, Xiangyu and Wang, Yuhao and Wang, Yejing and Zhang, Zijian and Sun, Yuqi and Li, Xiang and Wang, Maolin and Jia, Pengyue and Chen, Chong and others},
  booktitle={Proceedings of the 31st ACM SIGKDD Conference on Knowledge Discovery and Data Mining V. 2},
  pages={6096--6106},
  year={2025}
}

@article{wu2024survey,
  title={A survey on large language models for recommendation},
  author={Wu, Likang and Zheng, Zhi and Qiu, Zhaopeng and Wang, Hao and Gu, Hongchao and Shen, Tingjia and Qin, Chuan and Zhu, Chen and Zhu, Hengshu and Liu, Qi and others},
  journal={World Wide Web},
  volume={27},
  number={5},
  pages={60},
  year={2024},
  publisher={Springer}
}

@inproceedings{dai2023uncovering,
  title={Uncovering chatgpt’s capabilities in recommender systems},
  author={Dai, Sunhao and Shao, Ninglu and Zhao, Haiyuan and Yu, Weijie and Si, Zihua and Xu, Chen and Sun, Zhongxiang and Zhang, Xiao and Xu, Jun},
  booktitle={Proceedings of the 17th ACM Conference on Recommender Systems},
  pages={1126--1132},
  year={2023}
}

@article{zhao2024recommender,
  title={Recommender systems in the era of large language models (llms)},
  author={Zhao, Zihuai and Fan, Wenqi and Li, Jiatong and Liu, Yunqing and Mei, Xiaowei and Wang, Yiqi and Wen, Zhen and Wang, Fei and Zhao, Xiangyu and Tang, Jiliang and others},
  journal={IEEE Transactions on Knowledge and Data Engineering},
  volume={36},
  number={11},
  pages={6889--6907},
  year={2024},
  publisher={IEEE}
}

@inproceedings{lian2025egrec,
  title={EGRec: Leveraging Generative Rich Intents for Enhanced Recommendation with Large Language Models},
  author={Lian, Zhaorui and Geng, Binzong and Chang, Xiyu and Zhang, Yu and Ding, Ke and Lyu, Ziyu and Yuan, Guanghu and Li, Chengming and Yang, Min and Huan, Zhaoxin and others},
  booktitle={Companion Proceedings of the ACM on Web Conference 2025},
  pages={1113--1117},
  year={2025}
}

@article{shinn2023reflexion, 
  title={Reflexion: Language agents with verbal reinforcement learning},
  author={Shinn, Noah and Cassano, Federico and Gopinath, Ashwin and Narasimhan, Karthik and Yao, Shunyu},
  journal={Advances in Neural Information Processing Systems},
  volume={36},
  pages={8634--8652},
  year={2023}
}

@inproceedings{kweon2025uncertainty,
  title={Uncertainty Quantification and Decomposition for LLM-based Recommendation},
  author={Kweon, Wonbin and Jang, Sanghwan and Kang, SeongKu and Yu, Hwanjo},
  booktitle={Proceedings of the ACM on Web Conference 2025},
  pages={4889--4901},
  year={2025}
}

@article{wang2021survey,
  title={A survey on session-based recommender systems},
  author={Wang, Shoujin and Cao, Longbing and Wang, Yan and Sheng, Quan Z and Orgun, Mehmet A and Lian, Defu},
  journal={ACM Computing Surveys (CSUR)},
  volume={54},
  number={7},
  pages={1--38},
  year={2021},
  publisher={ACM New York, NY, USA}
}

@article{qiu2020exploiting,
  title={Exploiting cross-session information for session-based recommendation with graph neural networks},
  author={Qiu, Ruihong and Huang, Zi and Li, Jingjing and Yin, Hongzhi},
  journal={ACM Transactions on Information Systems (TOIS)},
  volume={38},
  number={3},
  pages={1--23},
  year={2020},
  publisher={ACM New York, NY, USA}
}

@inproceedings{ferrato2023challenges,
  title={Challenges for anonymous session-based recommender systems in indoor environments},
  author={Ferrato, Alessio},
  booktitle={Proceedings of the 17th ACM Conference on Recommender Systems},
  pages={1339--1341},
  year={2023}
}

@inproceedings{liu2020keywords,
  title={Keywords generation improves e-commerce session-based recommendation},
  author={Liu, Yuanxing and Ren, Zhaochun and Zhang, Wei-Nan and Che, Wanxiang and Liu, Ting and Yin, Dawei},
  booktitle={Proceedings of The Web Conference 2020},
  pages={1604--1614},
  year={2020}
}

@inproceedings{kersbergen2022serenade,
  title={Serenade-low-latency session-based recommendation in e-commerce at scale},
  author={Kersbergen, Barrie and Sprangers, Olivier and Schelter, Sebastian},
  booktitle={Proceedings of the 2022 International Conference on Management of Data},
  pages={150--159},
  year={2022}
}

@inproceedings{wang2025re2llm,
  title={Re2llm: Reflective reinforcement large language model for session-based recommendation},
  author={Wang, Ziyan and Du, Yingpeng and Sun, Zhu and Chua, Haoyan and Feng, Kaidong and Wang, Wenya and Zhang, Jie},
  booktitle={Proceedings of the AAAI Conference on Artificial Intelligence},
  volume={39},
  number={12},
  pages={12827--12835},
  year={2025}
}

@inproceedings{liu2018stamp,
  title={STAMP: short-term attention/memory priority model for session-based recommendation},
  author={Liu, Qiao and Zeng, Yifu and Mokhosi, Refuoe and Zhang, Haibin},
  booktitle={Proceedings of the 24th ACM SIGKDD international conference on knowledge discovery \& data mining},
  pages={1831--1839},
  year={2018}
}

@inproceedings{PO4ISR,
  title={Large language models for intent-driven session recommendations},
  author={Sun, Zhu and Liu, Hongyang and Qu, Xinghua and Feng, Kaidong and Wang, Yan and Ong, Yew Soon},
  booktitle={Proceedings of the 47th International ACM SIGIR Conference on Research and Development in Information Retrieval},
  pages={324--334},
  year={2024}
}

@article{LLM4SBR,
  title={Multi-view Intent Learning and Alignment with Large Language Models for Session-based Recommendation},
  author={Qiao, Shutong and Zhou, Wei and Wen, Junhao and Gao, Chen and Luo, Qun and Chen, Peixuan and Li, Yong},
  journal={ACM Transactions on Information Systems},
  volume={43},
  number={4},
  pages={1--25},
  year={2025},
  publisher={ACM New York, NY}
}

@inproceedings{RLMRec,
  title={Representation learning with large language models for recommendation},
  author={Ren, Xubin and Wei, Wei and Xia, Lianghao and Su, Lixin and Cheng, Suqi and Wang, Junfeng and Yin, Dawei and Huang, Chao},
  booktitle={Proceedings of the ACM Web Conference 2024},
  pages={3464--3475},
  year={2024}
}

@inproceedings{ProxySR,
  title={Unsupervised proxy selection for session-based recommender systems},
  author={Cho, Junsu and Kang, SeongKu and Hyun, Dongmin and Yu, Hwanjo},
  booktitle={Proceedings of the 44th International ACM SIGIR Conference on research and development in information retrieval},
  pages={327--336},
  year={2021}
}

@inproceedings{chen2022intent,
  title={Intent contrastive learning for sequential recommendation},
  author={Chen, Yongjun and Liu, Zhiwei and Li, Jia and McAuley, Julian and Xiong, Caiming},
  booktitle={Proceedings of the ACM web conference 2022},
  pages={2172--2182},
  year={2022}
}

@inproceedings{wang2021learning,
  title={Learning intents behind interactions with knowledge graph for recommendation},
  author={Wang, Xiang and Huang, Tinglin and Wang, Dingxian and Yuan, Yancheng and Liu, Zhenguang and He, Xiangnan and Chua, Tat-Seng},
  booktitle={Proceedings of the web conference 2021},
  pages={878--887},
  year={2021}
}

@inproceedings{yang2023based,
author = {Yang, Wei and Huo, Tengfei and Liu, Zhiqiang and Lu, Chi},
title = {Review-based Multi-intention Contrastive Learning for Recommendation},
year = {2023},
isbn = {9781450394086},
publisher = {Association for Computing Machinery},
address = {New York, NY, USA},
url = {https://doi.org/10.1145/3539618.3592053},
doi = {10.1145/3539618.3592053},
booktitle = {Proceedings of the 46th International ACM SIGIR Conference on Research and Development in Information Retrieval},
pages = {2339–2343},
numpages = {5},
location = {Taipei, Taiwan},
series = {SIGIR '23}
}

@inproceedings{li2023multi,
  title={Multi-intention oriented contrastive learning for sequential recommendation},
  author={Li, Xuewei and Sun, Aitong and Zhao, Mankun and Yu, Jian and Zhu, Kun and Jin, Di and Yu, Mei and Yu, Ruiguo},
  booktitle={Proceedings of the sixteenth ACM international conference on web search and data mining},
  pages={411--419},
  year={2023}
}

@inproceedings{qin2024intent,
  title={Intent contrastive learning with cross subsequences for sequential recommendation},
  author={Qin, Xiuyuan and Yuan, Huanhuan and Zhao, Pengpeng and Liu, Guanfeng and Zhuang, Fuzhen and Sheng, Victor S},
  booktitle={Proceedings of the 17th ACM international conference on web search and data mining},
  pages={548--556},
  year={2024}
}

@article{LLMESR,
  title={Llm-esr: Large language models enhancement for long-tailed sequential recommendation},
  author={Liu, Qidong and Wu, Xian and Wang, Yejing and Zhang, Zijian and Tian, Feng and Zheng, Yefeng and Zhao, Xiangyu},
  journal={Advances in Neural Information Processing Systems},
  volume={37},
  pages={26701--26727},
  year={2024}
}

@article{ELCRec,
  title={End-to-end learnable clustering for intent learning in recommendation},
  author={Liu, Yue and Zhu, Shihao and Xia, Jun and Ma, Yingwei and Ma, Jian and Liu, Xinwang and Yu, Shengju and Zhang, Kejun and Zhong, Wenliang},
  journal={Advances in Neural Information Processing Systems},
  volume={37},
  pages={5913--5949},
  year={2024}
}

@article{FAPAT,
  title={Enhancing user intent capture in session-based recommendation with attribute patterns},
  author={Liu, Xin and Li, Zheng and Gao, Yifan and Yang, Jingfeng and Cao, Tianyu and Wang, Zhengyang and Yin, Bing and Song, Yangqiu},
  journal={Advances in Neural Information Processing Systems},
  volume={36},
  pages={30821--30839},
  year={2023}
}

@inproceedings{MiasRec,
  title={Multi-intent-aware session-based recommendation},
  author={Choi, Minjin and Kim, Hye-young and Cho, Hyunsouk and Lee, Jongwuk},
  booktitle={Proceedings of the 47th international ACM SIGIR conference on research and development in information retrieval},
  pages={2532--2536},
  year={2024}
}

@inproceedings{CAFE,
  title={Coarse-to-fine sparse sequential recommendation},
  author={Li, Jiacheng and Zhao, Tong and Li, Jin and Chan, Jim and Faloutsos, Christos and Karypis, George and Pantel, Soo-Min and McAuley, Julian},
  booktitle={Proceedings of the 45th international ACM SIGIR conference on research and development in information retrieval},
  pages={2082--2086},
  year={2022}
}

@inproceedings{GCE-GNN,
  title={Global context enhanced graph neural networks for session-based recommendation},
  author={Wang, Ziyang and Wei, Wei and Cong, Gao and Li, Xiao-Li and Mao, Xian-Ling and Qiu, Minghui},
  booktitle={Proceedings of the 43rd international ACM SIGIR conference on research and development in information retrieval},
  pages={169--178},
  year={2020}
}

@inproceedings{GC-SAN,
  title={Graph contextualized self-attention network for session-based recommendation.},
  author={Xu, Chengfeng and Zhao, Pengpeng and Liu, Yanchi and Sheng, Victor S and Xu, Jiajie and Zhuang, Fuzhen and Fang, Junhua and Zhou, Xiaofang},
  booktitle={IJCAI},
  volume={19},
  number={2019},
  pages={3940--3946},
  year={2019}
}

@inproceedings{CoSAN,
  title={Collaborative self-attention network for session-based recommendation.},
  author={Luo, Anjing and Zhao, Pengpeng and Liu, Yanchi and Zhuang, Fuzhen and Wang, Deqing and Xu, Jiajie and Fang, Junhua and Sheng, Victor S},
  booktitle={IJCAI},
  pages={2591--2597},
  year={2020}
}

@inproceedings{SR-GNN,
  title={Session-based recommendation with graph neural networks},
  author={Wu, Shu and Tang, Yuyuan and Zhu, Yanqiao and Wang, Liang and Xie, Xing and Tan, Tieniu},
  booktitle={Proceedings of the AAAI conference on artificial intelligence},
  volume={33},
  number={01},
  pages={346--353},
  year={2019}
}

@inproceedings{narm,
  title={Neural attentive session-based recommendation},
  author={Li, Jing and Ren, Pengjie and Chen, Zhumin and Ren, Zhaochun and Lian, Tao and Ma, Jun},
  booktitle={Proceedings of the 2017 ACM on Conference on Information and Knowledge Management},
  pages={1419--1428},
  year={2017}
}

@article{gru4rec,
  title={Session-based Recommendations with Recurrent Neural Networks},
  author={Hidasi, B},
  journal={arXiv preprint arXiv:1511.06939},
  year={2015}
}

@article{transformer,
  title={Attention is all you need},
  author={Vaswani, A},
  journal={Advances in Neural Information Processing Systems},
  year={2017}
}

@inproceedings{bert4rec,
  title={BERT4Rec: Sequential recommendation with bidirectional encoder representations from transformer},
  author={Sun, Fei and Liu, Jun and Wu, Jian and Pei, Changhua and Lin, Xiao and Ou, Wenwu and Jiang, Peng},
  booktitle={Proceedings of the 28th ACM international conference on information and knowledge management},
  pages={1441--1450},
  year={2019}
}

@inproceedings{sasrec,
  title={Self-attentive sequential recommendation},
  author={Kang, Wang-Cheng and McAuley, Julian},
  booktitle={2018 IEEE international conference on data mining (ICDM)},
  pages={197--206},
  year={2018},
  organization={IEEE}
}

@inproceedings{selftrain1,
  title={Analyzing the effectiveness and applicability of co-training},
  author={Nigam, Kamal and Ghani, Rayid},
  booktitle={Proceedings of the ninth international conference on Information and knowledge management},
  pages={86--93},
  year={2000}
}

@article{selftrain2,
  title={Semi-supervised learning by entropy minimization},
  author={Grandvalet, Yves and Bengio, Yoshua},
  journal={Advances in neural information processing systems},
  volume={17},
  year={2004}
}
